\PassOptionsToPackage{prologue,dvipsnames}{xcolor}
\documentclass{article} 
\usepackage{iclr2025_conference,times}

\usepackage{hyperref}

\newcommand{\x}{\mathbf{x}}
\newcommand{\y}{\mathbf{y}}
\newcommand{\z}{\mathbf{z}}

\usepackage[dvipsnames]{xcolor} 

\hypersetup{
  breaklinks,
  colorlinks,
  linkcolor=BlueViolet,
  citecolor=BlueViolet,
  urlcolor=BlueViolet,
}

\newcommand\numeq[1]%
  {\stackrel{\scriptscriptstyle(\mkern-1.5mu#1\mkern-1.5mu)}{=}}

\usepackage{multirow} 
\usepackage{amssymb}
\usepackage{graphicx}
\usepackage[utf8]{inputenc} 
\usepackage[T1]{fontenc}    
\usepackage{hyperref}       
\usepackage{booktabs}       
\usepackage{amsfonts}       
\usepackage{nicefrac}       
\usepackage{microtype}      
\usepackage{algorithm}
\usepackage{algorithmic}
\usepackage{amsmath}
\usepackage{mathtools}
\usepackage[capitalize]{cleveref}
\usepackage{todonotes}
\usepackage{booktabs}
\usepackage{url}
\usepackage{xurl}

\usepackage{wrapfig}
\usepackage{enumitem}
\usepackage[small]{caption}
\newtheorem{prop}{Proposition}

\usepackage{amsmath,amsfonts,bm}

\def\eqref#1{equation~\ref{#1}}

\def\1{\bm{1}}

\DeclareMathAlphabet{\mathsfit}{\encodingdefault}{\sfdefault}{m}{sl}
\SetMathAlphabet{\mathsfit}{bold}{\encodingdefault}{\sfdefault}{bx}{n}

\title{Multi-Modal and Multi-Attribute Generation of Single Cells with CFGen}

\author{Alessandro Palma$^{1,2}$ \quad Till Richter$^{1,2}$ \quad Hanyi Zhang$^{1,2}$ \quad Manuel Lubetzki$^{1}$\\ \textbf{Alexander Tong$^{3,4}$ \quad Andrea Dittadi$^{1,2,5}$ \quad Fabian J. Theis$^{1,2}$\thanks{Correspondence to \url{fabian.theis@helmholtz-munich.de}} } 
\\
$^1$Helmholtz Munich \quad $^2$Technical University of Munich \quad $^3$Université de Montréal \\ $^4$Mila \quad $^5$MPI for Intelligent Systems, Tübingen}

\iclrfinalcopy 

\begin{document}

\maketitle

\begin{abstract}
Generative modeling of single-cell RNA-seq data is crucial for tasks like trajectory inference, batch effect removal, and simulation of realistic cellular data. However, recent deep generative models simulating synthetic single cells from noise operate on pre-processed continuous gene expression approximations, overlooking the discrete nature of single-cell data, which limits their effectiveness and hinders the incorporation of robust noise models. Additionally, aspects like controllable multi-modal and multi-label generation of cellular data remain underexplored. This work introduces CellFlow for Generation (CFGen), a flow-based conditional generative model that preserves the inherent discreteness of single-cell data. CFGen generates whole-genome multi-modal single-cell data reliably, improving the recovery of crucial biological data characteristics while tackling relevant generative tasks such as rare cell type augmentation and batch correction. We also introduce a novel framework for compositional data generation using Flow Matching. By showcasing CFGen on a diverse set of biological datasets and settings, we provide evidence of its value to the fields of computational biology and deep generative models.
\looseness=-1
\end{abstract}

\section{Introduction}
Single-cell transcriptomics has revolutionized our ability to study cell heterogeneity, revealing critical biological processes and cellular states \citep{RozenblattRosen2017}. Advances in single-cell RNA sequencing (scRNA-seq) enable high-throughput gene expression profiling across thousands of cells, providing valuable insights into cellular differentiation~\citep{Gulati2020}, disease progression~\citep{Zeng2019}, and responses to drug perturbations~\citep{ji2021machine}. Recognizing the complexity of a cell's molecular state, modern studies increasingly integrate additional measurements beyond gene expression, such as DNA accessibility \citep{grandi2022chromatin} to better characterize gene regulatory mechanisms~\citep{baysoy2023technological} or spatially resolved measurements to understand tissue organization~\citep{Marx2021}. Yet, technical bias and high experimental costs still hinder the homogeneous profiling of all possible cell states within the inspected biological process. Generative modeling offers a powerful approach to address these challenges by synthesizing biologically meaningful single-cell data, thereby uncovering underexplored cellular states and improving downstream analyses.
\looseness=-1

Generative models for single-cell data, in particular Variational Autoencoders (VAEs), have been extensively employed in representation learning~\citep{Lopez2018}, perturbation prediction~\citep{lotfollahi2019scgen, Lotfollahi2023, NEURIPS2022_aa933b5a} and trajectory inference \citep{Gayoso2023, Chen2022}. Recently, more complex approaches leveraging diffusion-based models~\citep{luo2024scdiffusion} or Generative Adversarial Networks (GAN)~\citep{marouf2020realistic} have paved the way for the task of synthetic data generation, demonstrating promising performance on realistic single-cell data modeling. Single-cell transcriptome data is inherently discrete, as gene expression is collected as the number of transcribed gene copies found experimentally. Due to the incompatibility of discrete data with continuous models such as Gaussian diffusion \citep{yang2023diffusion}, most approaches generate data pre-processed through normalization and scaling. This limits their flexibility to support downstream tasks centered around raw counts, such as batch correction \citep{Lopez2018}, differential gene expression \citep{Love2014, Chen2025, Heumos2024} and analyses where the total number of transcripts in a cell is meaningful \citep{Gulati2020}. Additionally, technical and biological effects in single-cell counts have been formalized under effective discrete noise models \citep{Hafemeister2019}, which should be incorporated into generative models for single-cell data to better approximate the underlying data generation process.
\looseness=-1

\begin{figure}
  \centering
  \includegraphics[width=1\textwidth]{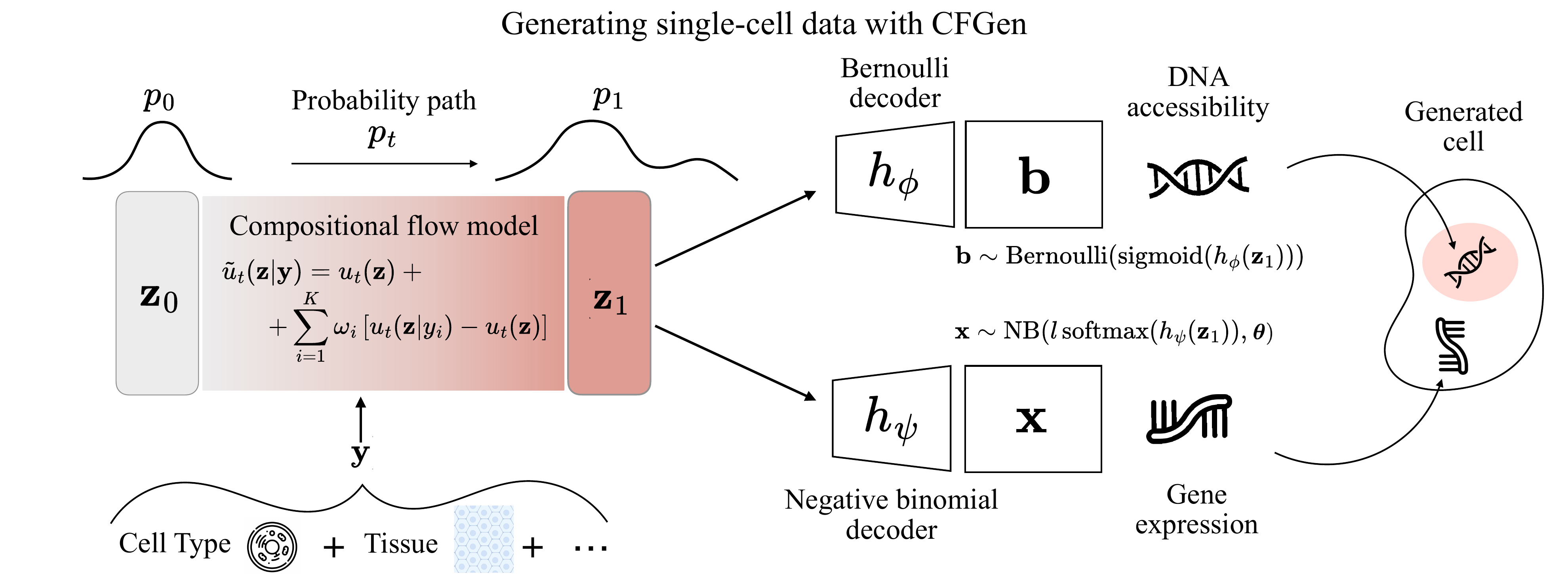}
  \caption{The CFGen generative model. A noise vector $\z_0$ sampled from a Gaussian prior $p_0$ is transformed into a latent cell representation $\z_1$ by a compositional flow, conditioned on multiple biological and technical attributes. Decoders for gene expression and DNA accessibility map $\z_1$ to the parameters of negative binomial and Bernoulli noise models, from which single-cell gene expression and DNA accessibility peaks are sampled.\looseness=-1}
  \label{fig: concept_fig}
\end{figure}

In this work, we present CellFlow for Generation (CFGen) (\cref{fig: concept_fig}), a conditional flow-based generative model designed to reproduce multi-modal single-cell discrete counts realistically. Our approach combines the expressiveness of recent Flow Matching techniques~\citep{albergo2022building, liu2022flow, lipman2022flow, dao2023flow, tong2023improving}
with modeling the statistical properties of single-cell data across multiple modalities, each following a distinct discrete likelihood model. Moreover, we extend the current literature on Flow Matching by introducing the concept of \textit{compositional guidance}, enabling the generation of cells conditioned on single attributes or combinations thereof in a controlled setting.\looseness=-1

We evaluate CFGen across multiple biological datasets, demonstrating its advantages in generative performance and downstream applications. Our main contributions are as follows:
\begin{itemize}[topsep=0pt,itemsep=0pt,partopsep=0pt,parsep=0.05cm,leftmargin=20pt]
    \item We introduce CFGen, a generative model for discrete multi-modal single-cell data that explicitly accounts for its key statistical properties under a specified noise model.
    \item We extend the Flow Matching framework to incorporate guidance for compositional generation under multiple attributes. 
    \item We show that our model's full-genome generative performance consistently outperforms existing single-cell generative models qualitatively and quantitatively on multiple biological datasets.
    \item We showcase the application of CFGen in enhancing downstream tasks, including robust data augmentation for improved classification of rare cell types and batch correction.
\end{itemize}
\section{Related work}
The synthetic generation of single-cell datasets is a well-established research direction pioneered by models using standard probabilistic methods to estimate gene-wise parameters in a single modality~\citep{zappia2017splatter, li2019statistical} or multiple modality setting~\citep{song2024scdesign3}. With the advent of deep generative models, VAE-based approaches have proven remarkably flexible, offering popular tools for batch correction \citep{Lopez2018}, modality integration~\citep{gayoso2021joint}, trajectory inference \citep{Gayoso2023} and perturbation prediction~\citep{lotfollahi2019scgen, bereket2023modelling}. Despite their relevance, most of the mentioned approaches focus on learning meaningful cellular representations or counterfactual predictions rather than generating synthetic datasets from noise. Such a task has instead been extensively explored by other works leveraging the expressive potential of diffusion models~\citep{luo2024scdiffusion, huang2024versatile}, Generative Adversarial Networks (GANs)~\citep{marouf2020realistic} and Large Language Models (LLMs) \citep{Levine2023} to produce realistic cells that approximate the observed data distribution. Our technical contribution builds upon Flow Matching \citep{albergo2022building, liu2022flow, lipman2022flow}, an efficient formulation of continuous normalizing flows for generative modeling. Since its introduction, Flow Matching has been successfully applied to optimal transport \citep{tong2023improving, eyring2023unbalancedness, pooladian2023multisample}, protein generation \citep{jing2024alphafold, yim2023fast}, interpolation on general geometries \citep{chen2023riemannian,  kapusniak2024metric} and guided conditional generation \citep{zheng2023guided}. Finally, Flow Matching showed promising performance in tasks involving scRNA-seq, such as learning cellular evolution across time~\citep{tong2023improving, kapusniak2024metric} and responses to drugs \citep{klein_genot}.
\looseness=-1

\section{Background}
\subsection{Deep generative modeling of single-cell data}
Single cells are represented as high-dimensional vectors of discrete counts, where each feature corresponds to a gene and its measurement reflects the number of transcripts detected in a cell. Technical bias and biological variation lead to unique characteristics in cells, including \textit{sparsity} and \textit{over-dispersion}. Sparsity arises from genes being inactive in specific cellular states (biological cause) or due to measurement dropouts in scRNA-seq (technical cause). Over-dispersion refers to the presence of greater variance than one would expect from a simple Poisson distribution of the count data (where the gene-wise variance equals the mean). This phenomenon is especially visible in highly expressed genes. Over-dispersed counts are typically modeled using a Negative Binomial (NB) distribution, parameterized by a mean $\mu$ and an inverse dispersion parameter $\theta$. Formally, given a nonnegative count expression matrix $\mathbf{X}\in\mathbb{N}_0^{N \times G}$ with $N$ cells and $G$ genes, entries $x_{ng}$ of the expression matrix are assumed to follow the negative binomial model:
\begin{equation}\label{eq: distr_nb}
    x_{ng} \sim \mathrm{NB}(\mu_{ng}, \theta_g) \: ,
\end{equation}
where \( \mu_{ng} \in \mathbb{R}_{\geq 0} \) is a cell-gene-specific mean and \( \theta_g \in \mathbb{R}_{> 0} \) is the gene-specific inverse dispersion. Thus, we assume each cell has an individual mean, while over-dispersion is modeled gene-wise. This parameterization of the negative binomial can be derived from a Poisson-gamma mixture, providing a natural formulation for the scRNA-seq likelihood (see \cref{sec: gamma_poisson_negative_binomial}).
\looseness=-1

When scRNA-seq is coupled with information on DNA accessibility, transcription measurements are complemented by a binary matrix $\mathbf{B}\in \{0,1\}^{N \times P}$, where $P$ is the number of DNA regions profiled for accessibility measured as the presence (1) or absence (0) of a signal peak. Here, each measurement independently follows the Bernoulli model $b_{np} \sim \mathrm{Bernoulli}(\pi_{np})$, with $\pi_{np}$ indicating a cell-gene-specific success probability. 

In most single-cell representation learning settings, a deep latent variable model is trained to map a latent space to the parameter space of the noise model via a decoder maximizing the log-likelihood of the data. Given a latent cell state $\z$, the likelihood parameters of each modality are inferred as 
\looseness=-1
\begin{align}
    \bm{\mu}=l  \bm{\rho}\:, 
    \quad
    \bm{\rho} = \mathrm{softmax}(h_\psi(\z))\:,
    \quad
    \bm{\pi} = \mathrm{sigmoid}(h_\phi(\z))\:,\label{eq: mu_model_2}
\end{align}
where $h_\psi$ and $h_\phi$ are modality-specific decoders and $l$ is the size factor, defined as the total number of counts of the generated cell. The vector $\bm{\rho}$ represents gene expression proportions.

\subsection{Continuous normalizing flows and Flow Matching}\label{sec: cnf_and_fm}
\textbf{Continuous Normalizing Flows (CNF).} \citet{chen2018neural} introduced CNFs as a generative model to approximate complex data distributions. Given data in a continuous domain $\mathcal{Z}\subset\mathbb{R}^d$, we define a time-dependent probability path $p:[0,1]\times \mathbb{R}^d \to \mathbb{R}_{\ge 0}$, transforming a tractable prior density $p_0$ into a more complex data density $p_1$, where we indicate the probability path at time $t$ as $p_t : \mathbb{R}^d \to \mathbb{R}_{\ge 0}$ such that $\int p_t(\z) \, \textrm{d}\z=1$. The probability path is formally \textit{generated} by a time-dependent smooth vector field $u_t: \mathbb{R}^d \to \mathbb{R}^d$, with $t\in [0,1]$, satisfying the continuity equation $\frac{\partial p_t}{\partial t} =  - \nabla \cdot (p_t u_t)$. The field $u_t$ is the time-derivative of an invertible \textit{flow} $\phi_t: \mathbb{R}^d \to \mathbb{R}^d$ following the Ordinary Differential Equation (ODE) $\frac{\mathrm{d}}{\mathrm{d}t} \phi_t(\z_0) = u_t(\phi_t(\z_0))$, where $\phi_0(\z_0)=\z_0$ and $\z_0$ is sampled from $p_0$. The flow $\phi_t$ defines a push-forward transformation $p_t=[\phi_t]_*p_0$, transforming the prior $p_0$ into the data density $p_1$. In other words, learning the vector field $u_t$ that governs the flow allows transporting samples from $p_0$ to $p_1$ by solving the ODE.\looseness=-1

\vskip -0.1cm
\textbf{Flow Matching.} Assume the goal is to model a complex data distribution $q$ from a prior $p_0$ by learning a continuous normalizing flow. One can marginalize the probability path $p_t$ as $p_t(\z) = \int p_t(\z | \z_1) q(\z_1)\textrm{d}\z_1$, where $\z_1$ indicates a sample from the data distribution $q$ and $p_t(\cdot | \z_1)$ is a \textit{conditional probability path} transporting noise to $\z_1$ under the boundary conditions $p_0(\z|\z_1)=p_0(\z)$ and $p_1(\z|\z_1)\approx\delta(\z-\z_1)$. Here, $\delta$ denotes a Dirac delta measure, which places all probability mass at $\mathbf{z}_1$. 
Note that, at $t=1$, the marginal distribution $p_1$ approximates the data distribution $q$.
Following the continuity equation, $p_t(\z)$ is generated by the \textit{marginal velocity field} $u_t(\z)$ that satisfies 
\looseness=-1
\begin{equation}\label{eq: marginal_field}
    u_t(\z) = \int u_t(\z | \z_1) \,\frac{p_t(\z|\z_1)q(\z_1)}{p_t(\z)}\,\textrm{d}\z_1 ,
\end{equation} 
where $u_t(\z | \z_1)$ is called \textit{conditional vector field}. 
Directly regressing $u_t(\z)$ is intractable. However, \citet{lipman2022flow} show that minimizing the Flow Matching objective 
\begin{equation}\label{eq: loss}
    \mathcal{L}_{\textrm{FM}}(\xi) = \mathbb{E}_{t\sim\mathcal{U}[0,1], q(\z_1), p_t(\z|\z_1)}\!\left[||v_{t, \xi}(\z) - u_t(\z|\z_1) ||^2\right]
\end{equation} 
corresponds to learning to approximate the marginal vector field $u_t$ with the time-conditioned neural network $v_{t,\xi}$ with parameters $\xi$. Defining $p_t(\z|\z_1)=\mathcal{N}(\alpha_t\z_1, \sigma_t^2\textbf{I})$ with the functions $\alpha_t, \sigma_t$ controlling the noise schedule, $u_t(\z|\z_1)$ has a closed form, and \cref{eq: loss} is tractable (see \cref{sec: fm_gau_paths} for more details). We define such a formulation as \textit{Gaussian marginal paths}.

\vskip -0.1cm
\textbf{Classifier-Free Guidance (CFG).}  
One can \textit{guide} data generation on a condition $y$ by learning the conditional marginal field $u_t(\z|y)$ via a time-conditioned neural network $v_{t, \xi}(\z,y)$. Given a guidance strength hyperparameter $\omega \in \mathbb{R}$, \citet{zheng2023guided} show that generating data points following the vector field  
$ \Tilde{u}_t(\cdot| y) = (1-\omega)u_{t}(\cdot) + \omega u_{t}(\cdot|y) $  
approximates sampling from the distribution $\Tilde{q}(\z|y) \propto q(\z)^{1-\omega}q(\z | y)^\omega$, where $q(\mathbf{z})$ and $q(\mathbf{z}|y)$ are, respectively, the unconditional and conditional data distributions. The parameter $\omega$ controls the trade-off between diversity and adherence to the condition. This approach enables guidance by interpolating between conditional and unconditional vector fields, both learned jointly during training.\looseness=-1  
\section{CFGen}
Our objective is to define a latent Flow-Matching-based generative model for discrete single-cell data, where each cell is measured through gene expression and, potentially, DNA accessibility. Our model, CFGen, is flexible: It can handle single and multiple modalities. Moreover, it supports guiding generation conditioned on single or combinations of attributes without needing to train a different model for each. In what follows, we present the assumptions and generative process formulation in the uni-modal and multi-modal settings. We additionally illustrate our novel approach to compositional guidance.\looseness=-1
\subsection{Uni-modal and single-attribute generation}\label{sec: unimodal_single_att}
Let $\mathbf{X}\in\mathbb{N}_0^{N \times G}$ be a single-cell matrix where an observed single-cell vector is $\mathbf{x} \in \mathbb{N}_0^{G}$, with $N$ and $G$ being the number of cells and genes. Additionally, let $\mathbf{y}\in\mathbb{N}_0^{N}$ be a vector of categorical labels associated with each observation. We also define  $l=\sum_{g=1}^G x_g$ as the size factor of an individual cell $\mathbf{x}$. 
\looseness=-1 

\vskip -0.1cm
\textbf{The generative process.} When the technical bias is negligible, we define the standard CFGen setting as the following generative model:
\begin{equation}\label{eq: cell_flow_generative}
    p(\x, \z, l, y) = p(\x | \z, l)p(\z | y, l)p(l)p(y) \; ,
\end{equation}
where $\z$ is a continuous latent variable modeling the cell state, and we assumed that (1) $\x$ is independent of $y$ conditionally on $\z$ and $l$, and (2) $l$ is independent of $y$. While \cref{eq: cell_flow_generative} defines a standard generative process, the factorization remains flexible based on data properties. Although related to existing VAE-based single-cell generative models, our proposed factorization is novel. We detail the relationship between \cref{eq: cell_flow_generative} and existing generative models in \cref{sec: rel_gen_models} and \ref{eq: size_factor_cond}.

\vskip -0.1cm
\textbf{Modeling the distributions in \cref{eq: cell_flow_generative}.} Each factor of \cref{eq: cell_flow_generative} is modeled separately: $p(y)$ is a categorical distribution $\mathrm{Cat}(N_y, \bm{\pi}_y)$ where $N_y$ is the number of categories and $\bm{\pi}_y$ a vector of $N_y$ class probabilities, and $p(l) = \mathrm{LogNormal}(\mu_l, \sigma^2_l)$. The parameters of $p(y)$ and $p(l)$ can be learned as maximum likelihood estimates over the dataset (see \cref{sec: sampling_from_noise}). Given an attribute class $y$ and size factor $l$ sampled from the respective distributions, $p(\z |y,l)$ is approximated by a conditional continuous normalizing flow $\phi_t(\cdot\,|y,l)$, with $t \in [0,1]$, learned via Flow Matching with Gaussian marginal paths (see \cref{sec: cnf_and_fm}, \cref{sec: fm_gau_paths} and \cref{sec: flow_model}). Such a flow transports samples $\z_0 \sim \mathcal{N}(\mathbf{0}, \mathbf{I})$ to latent cell representations $\z=\z_1= \phi_1(\z_0|y,l)$. Let $\mathbf{z}_t=\phi_t(\z_0|y,l)$. The time-derivative of the flow is a parameterized velocity function $v_{t, \xi} (\z_t, y, l)$. Finally, $p(\x | \z, l)$ samples from a negative binomial distribution with mean parameterized by a decoder $h_\psi$ as in \cref{eq: mu_model_2} and inverse dispersion modeled by a global parameter $\bm{\theta}$. In practice, $h_\psi$ and $\bm{\theta}$ are optimized before training the flow, together with an encoder $f_\eta$ that maps the data to a latent space (more details in \cref{sec: encoder_model}). We outline the reasons for training the encoder and the flow separately in \cref{sec: separate_training}.
\looseness=-1

\vskip -0.1cm

\textbf{Sampling in practice.} To generate a cell using CFGen as illustrated in \cref{eq: cell_flow_generative}, we first sample a size factor \( l \) and a condition \( y \) (the latter to specify a class). We then integrate the parameterized vector field $v_{t, \xi}(\z_t, y, l)$ with $t\in[0,1]$, starting from $\z_0 \sim \mathcal{N}(\mathbf{0}, \mathbf{I})$. We then take the simulated $\z_1 = \phi_1(\z_0|y,l)$ at $t=1$ as our latent $\z$ in \cref{eq: cell_flow_generative}. Finally, we sample \( \x \sim \mathrm{NB}(l \, \mathrm{softmax}(h_\psi(\mathbf{z}_1)), \bm{\theta})\).

\vskip -0.1cm
\textbf{Size factor as a technical effect.} When $l$ is influenced by technical effect under a categorical covariate $c \in \{1, \ldots, C\}$, we reformulate \cref{eq: cell_flow_generative} as 
$p(\x, \z, l, y) = \frac{1}{C} \sum_c p(\x | \z, l)p(\z | y, l)p(l|c)p(y)p(c)$,
where we assume that $\z$ is independent of $c$ given $l$ (i.e., $l$ contains all necessary technical effect information to guide the flow), and $y$ is independent of $c$. The last assumption derives from our choice of $y$ as an attribute encoding biological identity preserved across technical batches.
\subsection{Multi-modal and single-attribute generation}\label{sec: multimodal_setting} 
Let $\mathbf{X}$ and $\mathbf{y}$ be defined as in \cref{sec: unimodal_single_att}. In the multi-modal setting, we have additional access to a binary matrix $\mathbf{B}\in \{0,1\}^{N \times P}$ representing DNA region accessibility, with $P$ being the number of measured peaks. Each sample is, therefore, a tuple ($\x$, $\mathbf{b}$, $y$), where $\x$ and $\mathbf{b}$ are realizations of different discrete noise models (negative binomial and Bernoulli). Following \cref{eq: mu_model_2}, both parameters of the negative binomial and Bernoulli noise models are functions of the same latent variable $\z$, encoding a continuous cell state shared across modalities. We write the first factor in \cref{eq: cell_flow_generative} as 
\begin{equation}
    p(\x, \mathbf{b} | \z, l) \numeq{1} p(\x | \z, l)p(\mathbf{b} | \z, l) \ \numeq{2} p(\x | \z, l)p(\mathbf{b} | \z) \; ,
\end{equation}
where in (1) we use the fact that the likelihood of $\x$ and $\mathbf{b}$ are optimized disjointedly given $\z$, and in (2) that $\mathbf{b}$ is independent of the size factor $l$ (see \cref{eq: mu_model_2}). 
In simple terms, all the modalities are encoded to the same latent space used to train the conditional flow approximating $p(\z|y, l)$ in \cref{eq: cell_flow_generative} (more details in \cref{sec: encoder_model}). During generation, separate decoders $h_\psi$ and $h_\phi$ map a sampled latent variable $\z$ to the parameter spaces of the negative binomial and Bernoulli distributions, representing expression counts and binary DNA accessibility information, respectively (\cref{fig: concept_fig}). 

\subsection{Guided compositional generation with multiple attributes}\label{sec: multi_attr_guidance}
We extend CFG for Flow Matching \citep{zheng2023guided} to handle multiple attributes, enhancing control over the generative process in targeted data regions. This is especially relevant in scRNA-seq, where datasets are defined by several biological and technical covariates. Here, $\mathbf{Y} \in \mathbb{N}_0^{N \times K}$ represents a matrix of $K$ categorical attributes measured across $N$ cells. Rather than training separate models for each attribute combination, we compose multiple single-attribute flow models.

\vspace{-0.1cm}
Let $q(\z | \y)$ be the conditional data distribution, with $\y=(y_1,\ldots,y_ K)$ being a collection of observed categorical attributes. In analogy with CFG in diffusion models \citep{ho2022classifier}, we aim to implement a generative model to sample from $\Tilde{q}(\z|\y) \propto q(\z)\prod_{i=1}^{K} \left[\frac{q(\z|y_i)}{q(\z)}\right]^{\omega_i}$, where $\omega_i$ is the guidance strength for attribute $i$ (see \cref{sec: proof_prop}). 
Diffusion models generate data by learning to approximate the score of the time-dependent density, $\nabla_\mathbf{z} \log p_t(\mathbf{z}|\mathbf{y})$, with a neural network and using it to simulate a reverse diffusion Stochastic Differential Equation (SDE) transporting noise samples to generated data observations \citep{song2021scorebasedgenerativemodelingstochastic}. Importantly, the reverse diffusion SDE is associated with a deterministic \textit{probability flow ODE} with the same time-marginal densities \citep{yang2023diffusion}. 
\looseness=-1

\vspace{-0.1cm}
CFG in diffusion models can be used to generate data compositionally from different attribute classes. More specifically, \citet{liu2022compositional} demonstrated that compositional CFG is achievable through parameterizing the drift of the generating reverse SDE with the \textit{compositional score}:
\begin{equation}\label{eq: comp_gudance_score}
    \nabla_{\z} \log\Tilde{p}_t(\z|\y) = \nabla_{\z} \log p_t(\z) + \sum\nolimits_{i=1}^{K}\omega_i[\nabla_{\z} \log p_t(\z | y_i) - \nabla_{\z} \log p_t(\z)] \: .
\end{equation}
Following the direct relationship between Flow Matching and CFG provided in \citet{ho2022classifier}, we build the Flow Matching counterpart to \cref{eq: comp_gudance_score}. 

Provided that we have access to Flow Matching models for the unconditional marginal vector field $u_t(\z)$ and the single-attribute conditional fields $\{u_t(\z | y_i)\}_{i=1}^K$, the following holds:
\looseness=-1

\begin{prop}\label{prop: prop_1}
If the attributes $y_1,...,y_K$ are conditionally independent given $\z$, the vector field 
\begin{equation}\label{eq: comp_gudance_vec_field}
    \Tilde{u}_t(\z | \y) = u_t(\z) +  \sum\nolimits_{i=1}^K \omega_i[u_t(\z|y_i)-u_t(\z)]
\end{equation}
coincides with the velocity of the probability-flow ODE associated with the generative SDE of a diffusion model with a compositional score as in \cref{eq: comp_gudance_score}. 
\end{prop}
\vspace{-0.1cm}

We provide a proof for \cref{prop: prop_1} in \cref{sec: proof_prop}. In other words, the reversed diffusion SDE from compositional CFG admits a deterministic probability flow ODE with velocity as in \cref{eq: comp_gudance_vec_field}. Consequently, CFG sampling from compositions of attributes is obtained by integrating the parameterized field $\Tilde{v}_{t, \xi}(\z, \y) = v_{t, \xi}(\z) +  \sum_{i=1}^K \omega_i[v_{t, \xi}(\z,y_i)-v_{t, \xi}(\z)]$ starting from samples from a Gaussian prior $p_0$. Note that both conditional and unconditional fields are parameterized by the same model, which is learned by providing single-attribute conditioning with a certain probability during training (Algorithms \ref{alg:guided_flows_tr} and \ref{alg:sampling_guided_flows}). \looseness=-1
\section{Experiments}
In this section, we compare CFGen with existing models in uni-modal (\cref{sec: comparison_existing_methods}) and multi-modal (\cref{sec: results - multimodal}) generation across five datasets. We evaluate quantitatively by measuring distributional proximity between real and generated cells, and qualitatively by assessing how well models capture real data properties. In \cref{sec: results - multiattribute}, we show the effectiveness of multi-attribute generation in guiding synthetic samples towards specific biological labels and donors. Lastly, in Sections \ref{sec: results - augmentation} and \ref{sec: batch_correction}, we demonstrate that CFGen enhances rare cell type classification through targeted data augmentation and performs batch correction on par with widely used VAE-based models.
\looseness=-1

\subsection{Comparison with existing methods on uni-modal scRNA-seq generation}\label{sec: comparison_existing_methods}

We evaluate the performance of CFGen conditionally and unconditionally against three baselines.\looseness=-1

\begin{table}
\centering
\caption{Quantitative performance comparison of CFGen with conditional and unconditional single-cell generative models. Evaluation is performed based on distribution matching metrics (RBF-kernel MMD and 2-Wasserstein distance). Results are averaged across datasets generated using three different seeds.}
\label{tab:performance_comparison}
\resizebox{0.95\textwidth}{!}{%
\begin{tabular}{ccccccccc}
\hline
                   & \multicolumn{2}{c}{PBMC3K}  & \multicolumn{2}{c}{Dentate gyrus}  & \multicolumn{2}{c}{Tabula Muris} & \multicolumn{2}{c}{HLCA}  \\ \hline
                   & MMD ($\downarrow$)  & WD  ($\downarrow$)  & MMD  ($\downarrow$)  & WD   ($\downarrow$)  & MMD  ($\downarrow$)  & WD   ($\downarrow$)  & MMD  ($\downarrow$)  & WD  ($\downarrow$)  \\ \cline{2-9} 
                   & \multicolumn{8}{c}{Conditional}  \\ \hline
c-CFGen   & \textbf{0.85 \scriptsize{± 0.05}} & \textbf{16.94 \scriptsize{± 0.44}} & \textbf{1.12 \scriptsize{± 0.04}} & \textbf{21.55 \scriptsize{± 0.17}} & \textbf{0.19 \scriptsize{± 0.02}} & \textbf{7.39 \scriptsize{± 0.20}}  & \textbf{0.54 \scriptsize{± 0.02}} & \textbf{10.72 \scriptsize{± 0.08}}  \\
scDiffusion        & 1.27 \scriptsize{± 0.20} & 22.41 \scriptsize{± 1.21} & 1.22 \scriptsize{± 0.05} & 22.56 \scriptsize{± 0.10} & 0.24 \scriptsize{± 0.04} & 7.89 \scriptsize{± 0.45} & 0.96 \scriptsize{± 0.04} & 15.82 \scriptsize{± 0.45}  \\
scVI               & 0.94 \scriptsize{± 0.05} & 17.66 \scriptsize{± 0.29} & 1.15 \scriptsize{± 0.04} & 22.61 \scriptsize{± 0.23} & 0.26 \scriptsize{± 0.02} & 9.76 \scriptsize{± 0.53} & 0.58 \scriptsize{± 0.02} & 11.78 \scriptsize{± 0.19}  \\ \hline
                   & \multicolumn{8}{c}{Unconditional}  \\ \hline
u-CFGen & 0.44 \scriptsize{± 0.01} & 16.81 \scriptsize{± 0.06} & 0.42 \scriptsize{± 0.01} & \textbf{21.20 \scriptsize{± 0.02}} & \textbf{0.08 \scriptsize{± 0.00}} & \textbf{8.54 \scriptsize{± 0.06}} & \textbf{0.15 \scriptsize{± 0.01}} & \textbf{10.63 \scriptsize{± 0.01}}  \\
scGAN              & \textbf{0.36 \scriptsize{± 0.01}} & \textbf{15.54 \scriptsize{± 0.06}} & 0.42 \scriptsize{± 0.01} & 22.52 \scriptsize{± 0.03} & 0.25 \scriptsize{± 0.00} & 12.85 \scriptsize{± 0.04} & 0.18 \scriptsize{± 0.01} & 10.81 \scriptsize{± 0.01}  \\ \hline
\end{tabular}
}
\end{table}

\vskip -0.1cm
\textbf{Baselines.} We choose scVI~\citep{gayoso2021joint} and scDiffusion~\citep{luo2024scdiffusion} as conditional models and scGAN~\citep{marouf2020realistic} as unconditional baseline. The scVI model is based on a VAE architecture with a negative binomial decoder and performs generation by decoding low-dimensional Gaussian noise into parameters of the likelihood model. Conversely, scDiffusion and scGAN operate on a continuous-space domain, performing generation using standard latent diffusion \citep{Rombach_2022_CVPR} and GAN \citep{goodfellow2014generative} models. Thus, we train them using normalized counts (more in \cref{sec: baseline_descr}).

\vspace{-0.1cm}
\textbf{Datasets.} We assess model performance on four datasets of varying size: \textbf{(i)} PBMC3K\footnote{\url{https://satijalab.org/seurat/articles/pbmc3k_tutorial.html}} (2,638 cells from a healthy donor, clustering into 8 cell types), \textbf{(ii)} Dentate gyrus \citep{la2018rna} (18,213 cells from a developing mouse hippocampus), \textbf{(iii)} Tabula Muris \citep{tabula_muris2018} (245,389 cells from \textit{Mus musculus} across multiple tissues), and \textbf{(iv)} Human Lung Cell Atlas (HLCA)~\citep{sikkema2023integrated} (584,944 cells from 486 individuals across 49 datasets). Conditioning is performed on \textit{cell type} for all datasets except Tabula Muris, where we use the \textit{tissue} label. Dataset descriptions and pre-processing details are in \cref{preproc} and \cref{tab: dataset_specifics}.
\looseness=-1

\vspace{-0.1cm}
\textbf{Quantitative evaluation.} As evaluation metrics, we use distribution distances (RBF-kernel Maximum Mean Discrepancy (MMD)~\citep{borgwardt2006integrating} and 2-Wasserstein distance) computed between the Principal Component (PC) projections of generated and real test data in 30 dimensions. The generated data is embedded using the PC loadings of the real data for comparability. For conditional models, we evaluate the metrics per cell type and average the results. All evaluations are performed on a held-out set of cells, considering the whole genome, with a filtering step for low expression genes (see \cref{preproc}). 
\looseness=-1

\vspace{-0.1cm}
\textbf{Quantitative results.} In \cref{tab:performance_comparison}, we evaluate the generative performance of CFGen conditionally (c-CFGen) and unconditionally (u-CFGen) against the three baselines on the scRNA-seq generation task. CFGen consistently reaches the highest performance on conditional generation across biological categories and overcomes scGAN on three out of four datasets on the unconditional generation task. 

\begin{figure}[h]
\centering
\includegraphics[width=0.95\textwidth]{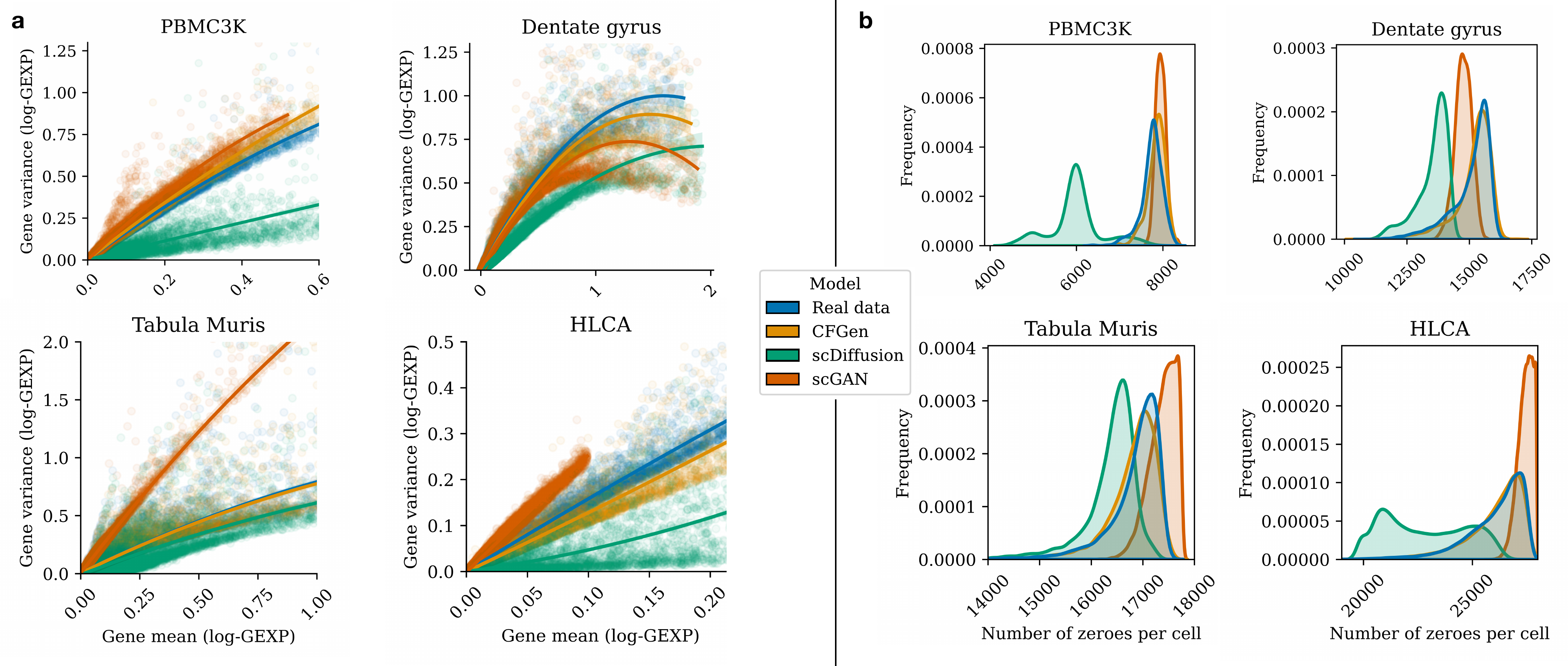}
\caption{\textbf{(a)} Comparison between the gene-wise empirical mean-variance trend in real data and samples from generative models. \textbf{(b)} Frequency of the number of zeroes per cell in real and generated data.\looseness=-1 }
\label{tab: scRNA_seq_properties}
\end{figure}
\vspace{-0.25cm}

\textbf{Qualitative evaluation.} Evaluating realistic data generation in biology requires more than distribution-matching metrics. We compare CFGen with diffusion and GAN-based single-cell generators on the task of modeling the probabilistic properties of the single-cell data. Specifically, we consider how well different methods recover the following aspects from real gene expression counts: (1) Sparsity: caused by technical biases in gene transcript detection or gene inactivity in specific contexts. (2) Over-dispersion: a nonlinear mean-variance relationship, modeled through the inverse dispersion parameter of a negative binomial distribution. (3) Discreteness.

\vspace{-0.1cm}
\textbf{Qualitative results.} In \cref{tab: scRNA_seq_properties}, we provide qualitative evidence that CFGen is more effective in recovering properties (1) and (2) compared to scDiffusion and scGAN, which assume a continuous data space. Property (3) naturally follows when modeling discrete counts with CFGen. Specifically, \cref{tab: scRNA_seq_properties}a shows that explicitly modeling counts with gene-specific inverse dispersion leads to better alignment of the generated gene-wise mean-variance relationship with real data. Additionally, \cref{tab: scRNA_seq_properties}b demonstrates the recovery of the actual distribution of zero counts per cell. In contrast, scDiffusion often shifts towards actively expressed genes, while scGAN tends to either under or overestimate data sparsity. Furthermore, CFGen is the only conditional model capable of generating plausible synthetic cells in terms of overlap with the real data distribution for large datasets such as the HLCA and Tabula Muris (see \cref{comp} and \cref{fig: comp_gen}). \looseness=-1 \looseness=-1

\subsection{Multi-modal generation}
\label{sec: results - multimodal}

We evaluate the qualitative and quantitative performance of CFGen at generating multi-modal data comprising gene expression and binary DNA-region accessibility. 

\vskip -0.1cm
\textbf{Baselines.} We compare CFGen with a VAE-based multi-modal generative model (MultiVI) \citep{ashuach2023multivi}.
For completeness, we add as baselines two single-modality generative models: PeakVI \citep{ashuach2022peakvi} (DNA accessibility) and scVI \citep{Lopez2018} (gene expression). Finally, we include scDiffusion and uni-modal CFGen (CFGen RNA) as baselines for scRNA-seq generation.\looseness=-1

\vskip -0.1cm
\textbf{Datasets.} We use the multiome PBMC10K dataset, made available by 10X Genomics \footnote{\url{https://www.10xgenomics.com/support/single-cell-multiome-atac-plus-gene-expression}}. Here, each cell is measured both in gene expression (RNA) and DNA accessibility (ATAC). The dataset consists of 10K cells across 25,604 genes and 40,086 peaks and was annotated with 15 cell types, with their respective marker peaks (enriched in accessible or inaccessible points) and genes.

\begin{wraptable}[12]{R}{0.5\textwidth}
\vskip -0.4cm
\caption{Comparison between CFGen, scDiffusion and VAE-based models on generating multiple single-cell modalities. We report distribution distance performance (RBF-kernel MMD and 2-Wasserstein distance) between real and generated cells across three seeds. Underlined values indicate the second-best performance.}
\vspace{-0.2cm}
\label{tab: multimodal_eval}
\centering
\resizebox{0.50\textwidth}{!}{%
\begin{tabular}{ccccc}
\hline
          & \multicolumn{2}{c}{RNA}                                           & \multicolumn{2}{c}{ATAC}                     \\ \hline
          & MMD ($\downarrow$)   & \multicolumn{1}{c|}{WD ($\downarrow$)}     & MMD ($\downarrow$)   & WD ($\downarrow$)     \\ \cline{2-5} 
CFGen multi.     & {\underline{0.89 \scriptsize{± 0.02}}}     & \multicolumn{1}{c|}{\textbf{13.90 \scriptsize{± 0.07}}} & \textbf{0.92 \scriptsize{± 0.02}} & \textbf{18.86 \scriptsize{± 0.37}} \\
CFGen RNA & \textbf{0.86 \scriptsize{± 0.02}} & \multicolumn{1}{c|}{\underline{14.30 \scriptsize{± 0.08}}}    & -                    & -                     \\
scDiff.   & 1.02 \scriptsize{± 0.02}          & \multicolumn{1}{c|}{14.82 \scriptsize{± 0.11}}          & -                    & -                     \\
MultiVI   & \textbf{0.86 \scriptsize{± 0.03}} & \multicolumn{1}{c|}{15.92 \scriptsize{± 0.25}}          & {\underline{ 0.96 \scriptsize{± 0.03}}}    & 21.09 \scriptsize{± 0.34}          \\
PeakVI    & -                    & \multicolumn{1}{c|}{-}                     & 1.49 \scriptsize{± 0.02}          & {\underline{20.84} \scriptsize{± 0.45}}    \\
scVI      & 0.95 \scriptsize{± 0.02}          & \multicolumn{1}{c|}{14.38 \scriptsize{± 0.11}}          & -                    & -                     \\ \hline
\end{tabular}
            
            }
\end{wraptable}

\vskip -0.1cm
\textbf{Evaluation.} We use the RBF-kernel MMD and 2-Wasserstein distances in the same setting described in \cref{sec: comparison_existing_methods}. Before comparison, we normalize both real and generated binary measurements of DNA accessibility via TF-IDF \citep{Aizawa2003} (in analogy to text mining). The metrics are computed in a 30-dimensional PC projection of the generated cells, using the PC loadings of the real data. RNA counts are treated as in \cref{sec: comparison_existing_methods}. In \cref{sec: additional_results_multimodal} we compare CFGen and MultiVI more biologically. Specifically, we assess how well they approximate per-cell-type marker peaks and gene expression (see  \cref{sec:multimodal_eval}). For each cell type, we compute the accessibility fraction and mean expression of literature-derived marker peaks and genes in both real and generated cells and report their correlation per cell type in \cref{fig: multimodal_figure}b.\looseness=-1

\vspace{-0.1cm}
\textbf{Results.} CFGen outperforms both MultiVI and PeakVI in modeling accessibility data based on distribution matching metrics (see \cref{tab: multimodal_eval}). When considering the RNA modality, our model surpasses scVI and scDiffusion in all metrics and MultiVI in terms of 2-Wasserstein distance. Qualitatively, \cref{fig: multimodal_figure}a shows substantial overlap between real and generated modalities. Finally, \cref{fig: multimodal_figure}b demonstrates that CFGen better approximates average marker peak accessibility and gene expression, outperforming MultiVI across all cell type categories.

\vspace{-0.3cm}

\subsection{Multi-attribute generation and guidance}
\label{sec: results - multiattribute}

We assess our approach to compositional guidance, as outlined in \cref{sec: multi_attr_guidance}. 
\looseness=-1

\vspace{-0.1cm}
\textbf{Datasets.} We showcase guidance on datasets with extensive technical variation, as one could combine different levels of biological and technical annotations to either augment rare cell type and batch combinations or control for the amount of technical effect added in simulation settings. Specifically, we consider \textbf{(i)} The NeurIPS 2021 dataset \citep{luecken2021sandbox} - 90,261 bone marrow cells from 12 healthy human donors. We use donor as a batch attribute and cell type as a biological covariate. We also consider \textbf{(ii)} the Tabular Muris dataset described in \cref{sec: comparison_existing_methods}, using tissue and Mouse ID as covariates. 
\looseness=-1

\vspace{-0.1cm}
\textbf{Evaluation.} The power of our guidance model is to generate data conditionally on an arbitrary subset of attributes---including unconditional generation---using a \textit{single trained model}. For each pair of covariates $(y_i,y_j)$, we evaluate generation on 500 generated cells varying the parameter $\omega_j$, keeping $\omega_i$ fixed. The expected result is conditional generation on $\omega_i$ when $\omega_j=0$ and generation from the intersection between the two attributes as $\omega_j$ increases. We additionally test the unconditional model, given by $\omega_i=\omega_j=0$, expected to recover the whole single-cell dataset. In the unconditional case, we generate as many cells as there are in the dataset to better evaluate the coverage. 

\begin{figure}[]
\includegraphics[width=0.9\textwidth]{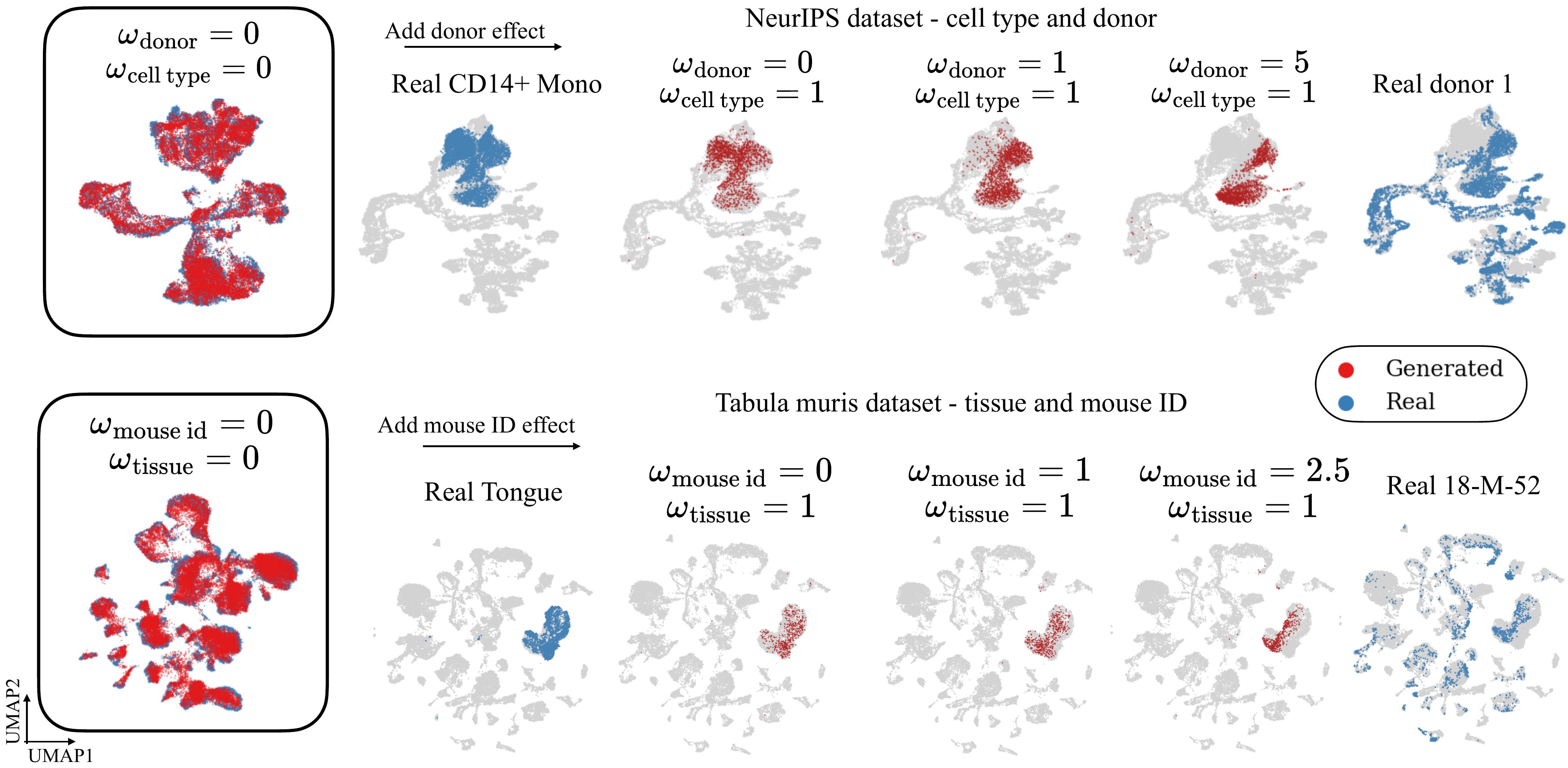}
\centering
\caption{Qualitative evaluation of guidance performance on attribute pairs in the NeurIPS 2021 and Tabula Muris datasets. Left: unconditional performance with guidance weights at 0. Moving right: simulate 500 cells, progressively increasing the guidance strength of one attribute while keeping the counterpart unchanged. }
\label{fig: multi_label_guidance}
\vspace{-0.7cm}
\end{figure}

\vspace{-0.1cm}
\textbf{Results.} Visual guidance results are shown in \cref{fig: multi_label_guidance}, with examples of double-attribute guidance between CD14+ monocytes and donor 1 for the NeurIPS 2021 dataset and tongue and mouse 18-M-52 for Tabula Muris. Unconditional generation recreates the original data (left-hand side) for both datasets. Setting guidance weights to zero for mouse ID and donor attributes leads to single-attribute conditional generation. Increasing the guidance weight steers the generation to the intersection of the two attributes. Quantitative results on attribute intersection generation quality are in \cref{sec: multilab_res}.\looseness=-1

\subsection{Application: data augmentation to improve classification of rare cell types}\label{sec: results - augmentation}
We explore using CFGen to improve cell-type classifier generalization by augmenting rare cell types in datasets. Previous work has shown data augmentation enhances cell type classification \citep{richter2024delineating}. As a classifier, we use scGPT~\citep{cui2024scgpt}, a transformer pre-trained on 33 million cells.\looseness=-1

\vskip -0.1cm
\textbf{Datasets.} We leverage two large datasets: \textbf{(i)} PBMC COVID \citep{yoshida2022local} - 422,220 blood cells from 93 patients ranging across paediatric and adult. \textbf{(ii)} The HLCA dataset described in \cref{sec: comparison_existing_methods}. Both datasets are processed by selecting 2000 highly variable features and holding out cells from 20\% of the donors.
\looseness=-1

\begin{wrapfigure}[15]{r}{0.50\textwidth}
  \centering
  \vspace{-0.60cm}
  \includegraphics[width=0.50\textwidth]{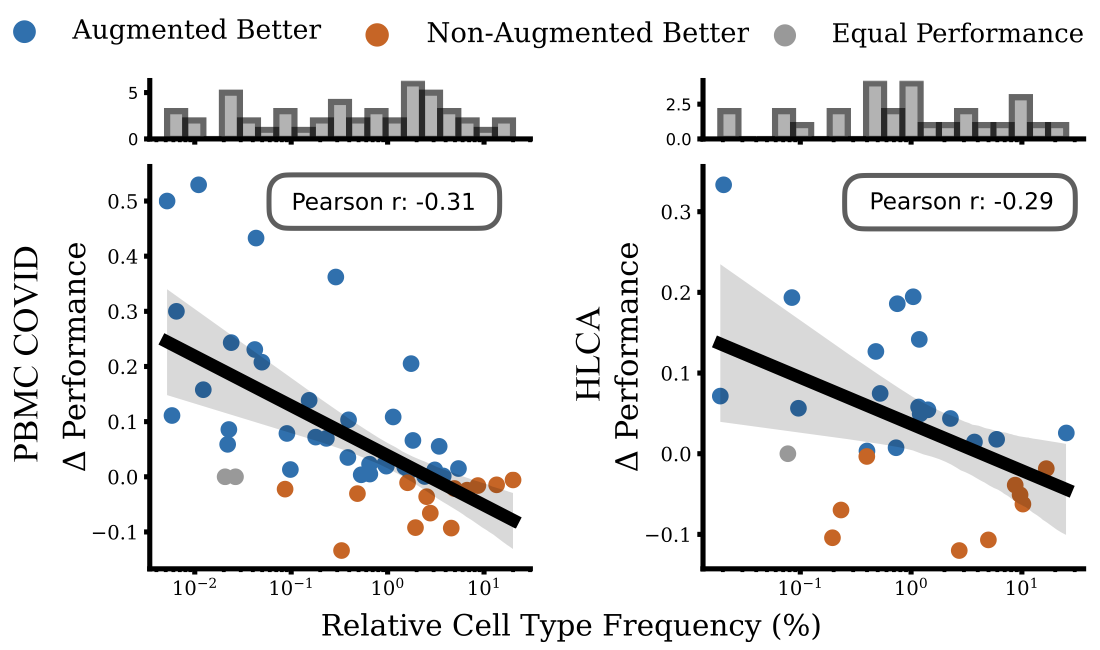}
\caption{Cell-type classification recall difference before and after augmentation as a function of cell type frequency. The classifier is a 10-nearest neighbor (kNN) model trained on the scGPT's representation space.}
  \label{fig: scgpt_exp}
\end{wrapfigure}

\vskip -0.1cm
\textbf{Evaluation.} We train CFGen on the PBMC COVID and HLCA training sets and successively augment both to 800,000 samples by upsampling rare cell types. For each cell type, we compute $\frac{1}{N_{\mathrm{ct}}}$, where $N_{\mathrm{ct}}$ is the total number of cells from a cell type \texttt{ct}. We then generate observations to fill the gap between the dataset size and 800,000 cells, sampling cells proportional to the inverse of their cell type frequency. This process yields significantly more observations for rare cell types. However, we still do not reach uniformity, as class imbalance may be biologically meaningful. Following the original publication, we train kNN cell-type classifiers on scGPT's embeddings from the original and augmented training sets, evaluating the recall performance on held-out donors. For each cell type, we assess if performance increases upon augmentation as a function of its frequency in the dataset.  
\looseness=-1

\vspace{-0.1cm}
\textbf{Results.} Our results are displayed in \cref{fig: scgpt_exp} for the two datasets. Remarkably, most cell types in the held-out dataset are better classified after augmentation, suggesting that CFGen not only generates reliable cell samples but can be a valuable supplement to relevant downstream tasks. Moreover, the performance difference between before and after augmentation is inversely proportional to the frequency of the cell type in the dataset. Therefore, the improvement in generalization is more accentuated for rare cell types. Finally, \cref{fig: augmentation_comp} in the Appendix shows that augmentation via CFGen outperforms the competing methods at improving the generalization performance on rare cell types in unseen donors. We provide raw cell type recall metrics in \cref{tab:ind_acc_pbmc} and \cref{tab:ind_acc_hlca}.
\looseness=-1

\subsection{Application: Batch correction}\label{sec: batch_correction}

\begin{figure}[]
\includegraphics[width=0.90\textwidth]{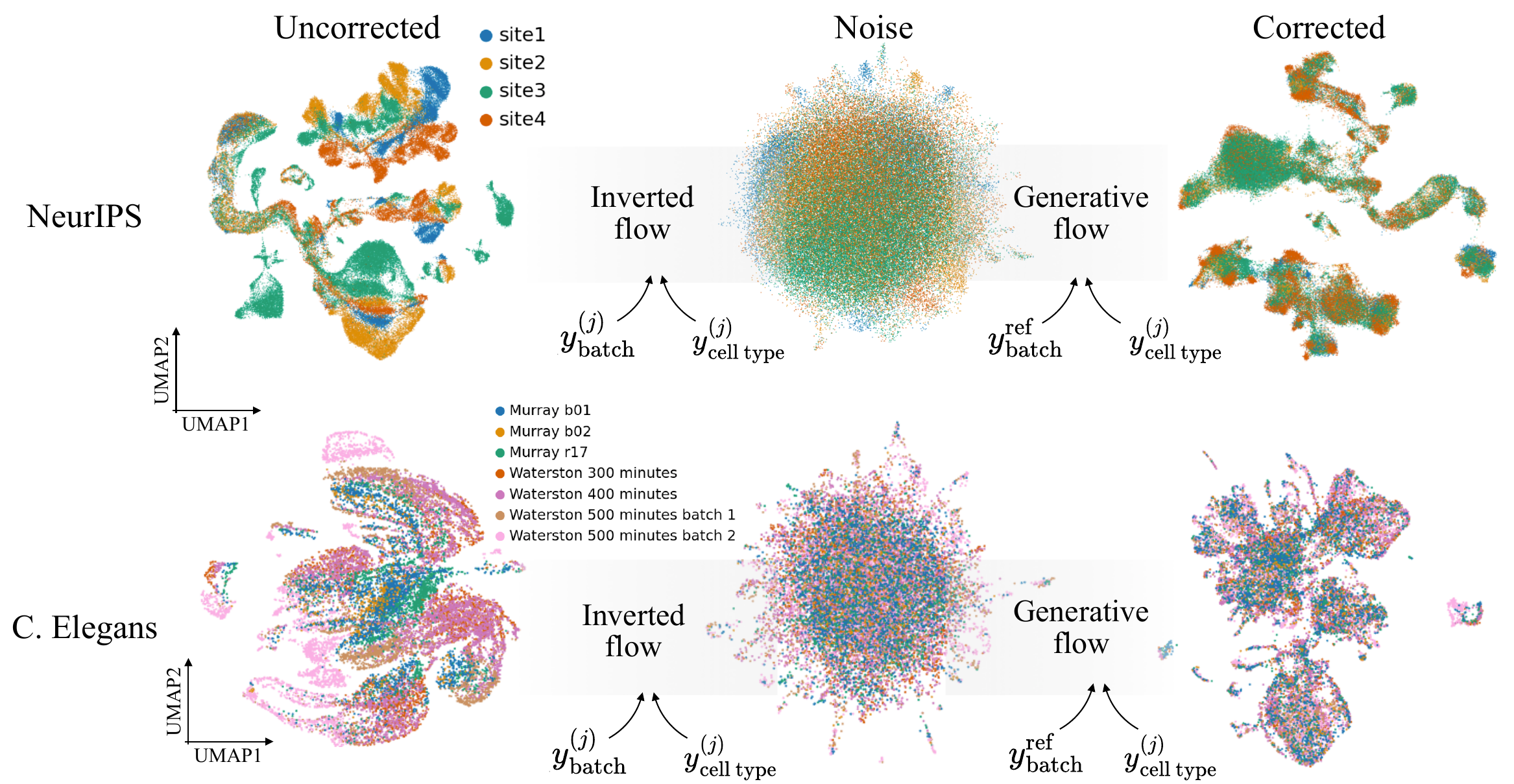}
\centering
\caption{To perform batch correction, the scRNA-seq latent distribution is mapped to the prior distribution by inverting the flow model. The resulting points are then transported back to the data domain based on a common reference batch label and the original cell type label to preserve the biological structure. Cells are colored by batch.\looseness=-1}
\label{fig: batch_correction}
\vskip -0.6 cm
\end{figure}

\begin{wraptable}[11]{R}{0.40\textwidth}
\centering
\caption{Average batch correction and biological conservation metrics from the scIB package comparing CFGen with VAE-based batch correction models in a 50-dimensional representation space. PC projections of the data are used to evaluate the uncorrected data.}
\label{tab: batch_correction_metrics}
\vspace{-0.2cm}
\resizebox{0.40\textwidth}{!}{%
\begin{tabular}{ccccc}
\hline
            & \multicolumn{2}{c}{NeurIPS}                & \multicolumn{2}{c}{C. Elegans}             \\ \hline
            & Batch ($\uparrow$) & Bio ($\uparrow$) & Batch ($\uparrow$) & Bio ($\uparrow$) \\ \cline{2-5} 
CFGen       & \textbf{0.63}    & 0.61                 & \textbf{0.68}    & \textbf{0.63}                    \\
scPoli      & 0.55             & 0.64                    & 0.61             & 0.56                   \\
scANVI      & 0.48             & \textbf{0.68}           & 0.61             & 0.59           \\
scVI        & 0.45             & 0.63                    & 0.58             & 0.55                   \\
Uncorrected & 0.33             & 0.62                    & 0.40             & 0.53                    \\ \hline
\end{tabular}}
\end{wraptable}

We apply multi-attribute CFGen to batch correction (see \cref{fig: batch_correction}), a common use case for generative models in scRNA-seq \citep{Tran2020, Luecken2021}. Given a dataset with batch labels, we choose a reference batch $y^{\mathrm{ref}}_{\mathrm{batch}}$. For a latent cell representation $\mathbf{z}_j$ with attributes $y^{(j)}_{\mathrm{batch}}$ and $y^{(j)}_{\mathrm{cell \, type}}$, we invert the generative flow to remove the attribute structure. Next, we simulate the forward flow from the obtained representations while fixing the cell type and assigning $y^{\mathrm{ref}}_{\mathrm{batch}}$ to all observations. Guidance weights regulate the preservation of biological versus batch labels.

\looseness=-1

\textbf{Datasets.} We evaluate CFGen as a batch correction method on two datasets: \textbf{(i)} The NeurIPS dataset described in \cref{sec: results - multiattribute}, using \textit{cell type} as a biological variable to preserve and \textit{acquisition site} as batch variable. \textbf{(ii)} The C. Elegans molecular atlas \citep{Packer2019}, which profiles 89,701 cells across 7 sources (batches). Similarly to (i), we use cell type as a biological annotation.  

\vskip -0.1cm
\textbf{Evaluation.} We compare our model with established VAE-based integration methods: scANVI \citep{Xu2021}, scVI, and scPoli \citep{DeDonno2023}. Using scIB metrics \citep{Luecken2021}, we assess batch correction and biological conservation based on neighborhood composition in the embedding space (see \cref{sec: batch_correction_appendix}). All methods are evaluated on a 50-dimensional latent space, with scores from the PC representation of uncorrected data included for comparison. We find that setting $\omega_{\mathrm{batch}}=1$ and $\omega_{\mathrm{cell \, type}}=2$ for C.Elegans and $\omega_{\mathrm{batch}}=2$ and $\omega_{\mathrm{cell \, type}}=1$ for NeurIPS preserves cell type variation while correcting for technical variation (see \cref{app:guidance_selection} for more details on the selection).

\vskip -0.1cm
\textbf{Results.} We present our technical effect correction approach alongside qualitative results in \cref{fig: batch_correction}, illustrating batch mixing performance across both datasets. In \cref{tab: batch_correction_metrics}, CFGen is benchmarked against baseline methods. Our model achieves superior batch correction, surpassing the second-best approach by $8\%$ on the NeurIPS dataset and $7\%$ on C. Elegans. On the latter dataset, CFGen additionally outperforms baseline models in preserving biological structure after correction. \looseness=-1
\section{Conclusion}
We presented CFGen, a conditional latent flow-based generative model for single-cell discrete data that combines state-of-the-art generative approaches with rigorous probabilistic considerations. CFGen incorporates established noise models to sample realistic gene expression and DNA accessibility states, with promising applications in data augmentation and batch correction tasks. Furthermore, our model demonstrates improved performance over existing generative frameworks, reproducing data more faithfully across modalities. Our core machine learning contribution extends classifier-free guidance in Flow Matching with compositional generation of multiple attributes. Overall, CFGen represents a significant advancement in the simulation and augmentation of single-cell data, offering the research community powerful tools to support biological analysis.

\vspace{-0.15cm}
\textbf{Limitations.} Our framework relies on multiple assumptions, including independence in the data, which may not hold in all biological contexts. Thus, exploring data characteristics is essential before using CFGen for generation. Furthermore, we currently train the autoencoder-based representation framework separately from the generative flow, which can be inefficient and memory-intensive. 

\newpage

\section*{Ethics statement}
This work explores the core features of single-cell data and examines how capturing complex, high-dimensional cellular information can assist in answering biological questions. We aim to release CFGen as a user-friendly, open-source tool to facilitate its adoption in single-cell analysis. Given its application in biological research, CFGen may be utilized in sensitive environments that involve clinical data and patient information.

\section*{Reproducibility statement}
Reproduction details are reported in the Appendix and main text. The proof for \cref{prop: prop_1} is extensively described in \cref{sec: proof_prop}, while prior knowledge on Flow Matching and classifier-free guidance is provided in \cref{sec: fm_gau_paths} and \cref{eq: score_flow_rel}. Algorithms for training and sampling with CFGen are reported in \cref{sec: algos}. We introduce a thorough model description of both autoencoder and flow components together with modeling choices in \cref{sec: model_details}. Baselines and their characteristics are reported in \cref{sec: baseline_descr}. All datasets are publicly available and their source publications are referenced in the main text. We additionally summarize dataset characteristics in \cref{tab: dataset_specifics}. Metrics and experimental setups are detailed in \cref{sec: metrics_and_experiments}. 

\section*{Acknowledgments}
A.P. and T.R. are supported by the Helmholtz Association under the joint research school Munich School for Data Science (MUDS). Additionally, A.P., T.R. and F.J.T. acknowledge support from the German Federal Ministry of Education and Research (BMBF) through grant numbers 031L0289A and 01IS18053A. T.R. and F.J.T. also acknowledge support from the Helmholtz Association’s Initiative and Networking Fund via the CausalCellDynamics project (grant number Interlabs-0029). Additionally, F.J.T. acknowledges support from the European Union (ERC, DeepCell - grant number 101054957). Finally, A.D. acknowledges support from G-Research.

\bibliography{iclr2025_conference}
\bibliographystyle{iclr2025_conference}

\clearpage
\appendix

\renewcommand{\thefigure}{A\arabic{figure}} 
\setcounter{figure}{0}

\section{Datasets and code}
We make the code for CFGen as well as the links to pre-processed datasets available at \url{https://github.com/theislab/CFGen}. 

\section{Theoretical supplement}
\subsection{Poisson-gamma and negative binomial distribution}\label{sec: gamma_poisson_negative_binomial}
A possible parameterization of the negative binomial distribution is via a mean $\mu$ and an inverse dispersion parameter $\theta$, with the Probability Mass Function (PMF):
\begin{equation}
p_{\textrm{NB}}(x \mid \mu, \theta) = \frac{\Gamma(\theta + x)}{x! \Gamma(\theta)}\Big(\frac{\theta}{\theta+\mu}\Big)^\theta\Big(\frac{\mu}{\theta+\mu}\Big)^x\: .
\end{equation}

where $\Gamma$ is the gamma function. One can show that the negative binomial distribution is obtained as a \textit{continuous mixture} of Poisson distributions with a gamma-distributed rate. More formally, define a Poisson model $x \sim \mathrm{Poisson}(\lambda)$ with $\lambda \geq 0$. The parameter $\lambda$ represents both the mean and variance of the distribution. Since the mean and variance are equal, the Poisson model is unsuitable for modeling over-dispersed counts (i.e., when the variance exceeds the mean). A way to overcome this limitation is to model the rate of the Poisson distribution as a random variable following a gamma distribution:
\begin{align}
    & x \sim \mathrm{Poisson}(\lambda) \: , \\  
    & \lambda \sim \mathrm{Gamma} \left( \theta, \frac{\mu}{\theta} \right) \: ,
\end{align}
where $\theta > 0$ is the shape parameter and $\frac{\mu}{\theta}$ is the scale parameter. Marginalizing out $\lambda$ in the PMF of the Poisson distribution recovers the PMF of a negative binomial with mean $\mu$ and inverse dispersion $\theta$.\looseness=-1

Crucially, the variance of this negative binomial parameterization is $\mu+\frac{\mu^2}{\theta}$. Since the additional variance term $\frac{\mu^2}{\theta}$ is always non-negative, the variance always exceeds the mean for finite $\theta$, making the negative binomial distribution well-suited for modeling over-dispersed count data.

\subsection{Flow Matching with Gaussian Paths}\label{sec: fm_gau_paths}
Flow Matching \citep{lipman2022flow} learns a time-dependent vector field $u_t(\z)$, where $t \in [0,1]$, generating a probability path $p_t(\z)$. Here, $p_0 = \mathcal{N}(\mathbf{0}, \mathbf{I})$ is the standard Gaussian prior, and $p_1$ represents a more complex target distribution. The marginal $p_t(\z)$ is commonly formulated as a mixture:
$$ p_t(\z) = \int p_t(\z|\z_1)q(\z_1)\, \mathrm{d}\z_1 ,$$
where $q$ denotes the target data distribution. The marginal vector field that generates such a mixture of paths is given by:
$$ u_t(\z) = \int u_t(\z | \z_1) \frac{p_t(\z|\z_1)q(\z_1)}{p_t(\z)}\, \textrm{d}\z_1 \; .$$
While the marginal vector field $u_t(\z)$ is intractable, the conditional vector field $u_t(\z|\z_1)$ has a closed-form expression when given an observed data point $\z_1$ and a pre-defined choice of the probability path $p_t(\z | \z_1)$. This path satisfies the boundary conditions $p_0(\cdot|\z_1)=p_0$ and $p_1(\cdot|\z_1)=\delta(\z-\z_1)$, where the Dirac delta measure $\delta(\z-\z_1)$ ensures that the distribution at time $t = 1$ is a point mass at the observed data point $\z_1$. Notably, regression on $u_t(\z|\z_1)$ admits the same minimizer as on $u_t(\z)$, making it suitable for use as a target during training.

Following \citet{lipman2022flow}, we assume Gaussian probability paths for the transformation at each time $t$, defined by:
$$ p_t(\z|\z_1) = \mathcal{N}(\z|\alpha_t\z_1, \sigma_t^2 \mathbf{I}) ,$$
where $\alpha_t$ and $\sigma_t$ are the scheduler parameters, with $\alpha_0=0$, $\sigma_1=0$, $\alpha_1=1$, and $\sigma_0=1$. In this work, we use standard linear scheduling, where $\alpha_t = t$ and $\sigma_t = 1 - t$, linearly interpolating between the initial and final values.

\subsection{The relationship between Flow Matching and classifier-free guidance (CFG)}\label{eq: score_flow_rel}
\citet{zheng2023guided} draw a relationship between CFG in score-based models \citep{ho2022classifier} and the Flow Matching vector field $u_t$. Specifically, the authors show that the following relationship between the score $\nabla_{\mathbf{z}} \log p_t(\mathbf{z}|y)$ and the marginal vector field $u_t(\mathbf{z}|y)$ holds:
\begin{equation}\label{eq: relation_score_flow}
    u_t(\z|y) = a_t \z + b_t \nabla_{\z} \log p_t(\z|y) \: ,
\end{equation}
both in the conditional case and when $y = \varnothing$, with $a_t = \frac{\dot{\alpha}_t}{\alpha_t}$ and $b_t = \left(\dot{\alpha}_t \sigma_t - \alpha_t \dot{\sigma}_t \right) \frac{\sigma_t}{\alpha_t}$. 

Let the equation
$$ \nabla_{\z} \log \Tilde{p}_t(\z|y) = (1-\omega) \nabla_{\z} \log p_t(\z) + \omega \nabla_{\z} \log p_t(\z|y) $$
be the CFG score as formulated by \citet{ho2022classifier} with guidance strength $\omega$. \citet{zheng2023guided} define the vector field of classifier-free Flow Matching as 
\begin{equation}\label{eq: class_free_fm}
    \Tilde{u}_t(\z|y) = (1-\omega) u_t(\z) + \omega \, u_t(\z|y) \: .
\end{equation}
This vector field is related to the CFG score by
$$ \Tilde{u}_t(\z|y) = a_t \z + b_t \nabla_{\z} \log \Tilde{p}_t(\z|y) \: ,$$
which is derived by substituting \cref{eq: relation_score_flow} into \cref{eq: class_free_fm}. 

\subsection{Proof of \cref{prop: prop_1}}\label{sec: proof_prop}

\setcounter{prop}{0}
\begin{prop}
If the attributes $y_1, \dots, y_K$ are conditionally independent given $\z$, the vector field 
\begin{equation}
    \Tilde{u}_t(\z | \y) = u_t(\z) +  \sum_{i=1}^K \omega_i \left[ u_t(\z|y_i) - u_t(\z) \right] \nonumber
\end{equation}
coincides with the velocity of the probability-flow ODE associated with the generative SDE of a diffusion model with the compositional score as in \cref{eq: comp_gudance_score}. 
\end{prop}

\textit{Proof. (\cref{prop: prop_1})} We first justify the conditional independence assumption and subsequently prove the equality.

\paragraph{Conditional independence assumption.} Given a variable $\z$ and a set of attributes $\y=y_1,...,y_K$, we define the attribute-conditioned marginal probability path $p_t(\z|y_1,...,y_K)$, with $t\in[0,1]$. Under the assumption that the attributes are conditionally independent given $\z$, one can rewrite the marginal path as follows:
\begin{equation}\label{eq: comp_score_joint}
    p_t(\z|y_1,...,y_K) \propto p_t(\z,y_1,...,y_K) = p_t(\z) \prod_{i=1}^K p_t(y_i | \z) \propto p_t(\z) \prod_{i=1}^K \frac{p_t(\z | y_i)}{p_t(\z)}\; .
\end{equation}
Taking the logarithm and then the gradient with respect to $\z$ on both sides in \cref{eq: comp_score_joint}, we obtain:
\begin{equation}\label{eq: score_decomposition}
    \nabla_{\z} \log p_t(\z|y_1,...,y_K) = \nabla_{\z} \log p_t(\z) + \sum_{i=1}^K \left[\nabla_{\z} \log p_t(\z|y_i) - \nabla_{\z} \log p_t(\z) \right] \; .
\end{equation}
In CFG the goal is to sample with attribute-specific guidance strengths $\{\omega_i\}_{i=1}^K$ according to a modified conditional data distribution
\begin{equation}
    \Tilde{q}(\z|y_1,...,y_K)\propto q(\z) \prod_{i=1}^K \left[\frac{q(\z | y_i)}{q(\z)} \right]^{\omega_i} \; .
\end{equation}
In terms of generative probabilistic paths, the score in \cref{eq: score_decomposition} becomes:
\begin{equation}\label{eq: score_decomposition_omega}
    \nabla_{\z} \log \Tilde{p}_t(\z|y_1,...,y_K) = \nabla_{\z} \log p_t(\z) + \sum_{i=1}^K \omega_i \left[\nabla_{\z} \log p_t(\z|y_i) - \nabla_{\z} \log p_t(\z) \right] \; .
\end{equation}
In score-based models, the formulation in \cref{eq: score_decomposition_omega} is used to parameterize the drift of the reverse-time SDE that generates data points conditionally on the attributes $y_1,...,y_K$ with guidance strengths $\{\omega_i\}_{i=1}^K$.\looseness=-1

\paragraph{Proof of equality.} Following the standard theory of score-based models \citep{yang2023diffusion} and their compositional extension \citep{liu2022compositional}, we first note that one can use the compositional CFG score  
\begin{equation}\label{eq: comp_guidance_score}
    \nabla_{\z} \log \Tilde{p}_t(\z|\y) = \nabla_{\z} \log p_t(\z) + \sum_{i=1}^K \omega_i \left[\nabla_{\z} \log p_t(\z|y_i) - \nabla_{\z} \log p_t(\z) \right] \; .
\end{equation}
to simulate the generative probability-flow ODE:  
\begin{equation}\label{eq: prob_flow_ode}
    \dot{\z} = f_t \z - \frac{1}{2}g_t^2 \nabla_\z \log \Tilde{p}_t(\z|\y) \; .
\end{equation}
Given a scheduling pair $(\alpha_t, \sigma_t)$, one can show \citep{kingma2021variational} that the drift and diffusion coefficients of a score-based diffusion model following the formulation from  \citet{song2021scorebasedgenerativemodelingstochastic} satisfy  
\begin{equation}
    f_t=\frac{\mathrm{d}\log \alpha_t}{\mathrm{d}t}, \quad g_t^2=\frac{\mathrm{d}\sigma_t^2}{\mathrm{d}t}-2\frac{\mathrm{d}\log \alpha_t}{\mathrm{d}t}\sigma_t^2 \:.
\end{equation}
From these, we derive  
\begin{equation}
    a_t = \frac{\dot{\alpha}_t}{\alpha_t}, \quad b_t = (\dot{\alpha}_t\sigma_t-\alpha_t\dot{\sigma}_t)\frac{\sigma_t}{\alpha_t} \:.
\end{equation}
Thus, rewriting \cref{eq: prob_flow_ode}, we obtain the probability-flow ODE in the form:  
\begin{equation}\label{eq: prob_flow_ode_final}
    \dot{\z} = a_t \z + b_t \nabla_\z \log \Tilde{p}_t(\z|\y) \; .
\end{equation}
Next, using results from \citet{zheng2023guided}, as discussed in \cref{eq: score_flow_rel}, we write the unconditional and conditional vector fields as  
\begin{align}
    &u_t(\z) = a_t \z + b_t \nabla_{\z} \log p_t(\z) \: , \label{eq: uncond_vec_field} \\ 
    &u_t(\z|y_i) = a_t \z + b_t \nabla_{\z} \log p_t(\z|y_i) \: . \label{eq: cond_vec_field}
\end{align}
Plugging \cref{eq: uncond_vec_field,eq: cond_vec_field} into the \cref{eq: comp_gudance_vec_field} for the compositional flow, we obtain  
\begin{align}
    \Tilde{u}_t(\z|\y) &= u_t(\z) + \sum_{i=1}^K \omega_i \left[u_t(\z|y_i) - u_t(\z)\right] \\
    &= a_t \z + b_t \left[\nabla_\z \log p_t(\z) + \sum_{i=1}^K \omega_i \left(\nabla_\z \log p_t(\z | y_i)-\nabla_\z \log p_t(\z) \right) \right] \\
    &= a_t \z + b_t \nabla_{\z} \log \Tilde{p}_t(\z|\mathbf{y}) \: .
\end{align}
This matches \cref{eq: prob_flow_ode_final}, completing the proof.

\subsection{Relationship with existing single-cell generative models}\label{sec: rel_gen_models}
Although likelihood-based models are standard in the single-cell literature, CFGen leverages a novel factorization scheme, as depicted in \cref{eq: cell_flow_generative}. Below, we outline key differences between our approach and standard single-cell VAEs:

\begin{itemize}
    \item In scVI \citep{Lopez2018}, the conditioning is applied during the decoding phase rather than at the prior level for the latent variable $p(\mathbf{z})$. In contrast, CFGen conditions directly on the latent prior $p(\mathbf{z}|y)$, allowing it to sample from multiple modes. This enables a more flexible generation compared to traditional approaches that rely solely on decoder-based conditioning.
    \item Consequently, the likelihood term $p(\mathbf{x}|\mathbf{z}, l)$ differs from that of most single-cell VAEs. In CFGen, this term is modeled using an unconditional decoder since the conditioning on $y$ is already incorporated in the flow-based generation of $\mathbf{z}$. Conversely, standard conditional VAEs must explicitly feed the label into both the encoder and decoder.
    \item While some VAEs incorporate a conditional prior \citep{Xu2021}, they are primarily designed for representation learning rather than generative modeling. These models often under-regularize the latent space to favor structure and reconstruction. In contrast, CFGen enforces a \textit{strong} conditional flow-based prior on the latent cell representation, avoiding the trade-off between Kullback-Leibler divergence minimization and likelihood optimization. As a result, it can generate high-quality samples without requiring a highly regularized latent space.\looseness=-1
    \item Although previous works have explored defining a distribution over library size, our approach uniquely integrates it into the generative process through our specific factorization in \cref{eq: cell_flow_generative}. Typically, the library size is used merely as a scaling factor in likelihood optimization, applied to the post-softmax output of the decoder. CFGen instead employs it as a conditioning variable for sampling from the flow-based conditional prior, ensuring a formally sound integration. Specifically, we define the library size as a conditioning attribute and factorize our latent variable model to generate $\mathbf{z}$ from $p(\mathbf{z}| y, l)$. To the best of our knowledge, this formulation has not been explored before. 
\end{itemize}

\subsection{Conditioning on the size factor}\label{eq: size_factor_cond}
Single-cell VAEs, like scVI, offer the option to learn a log-normal distribution over the size factor, which can then be sampled. This process is similar to our approach, as both methods fit a log-normal distribution over the library size in the data. However, the key distinction lies in how the size factor interacts with the latent cellular state.

In scVI, the latent cellular state and the size factor are sampled independently. In contrast, CFGen provides the option to bias the sampling by the size factor. More specifically, CFGen samples a latent state that inherently accounts for cell size, whereas scVI does not.

\begin{figure}[H]
\centering
\includegraphics[width=0.90\textwidth]{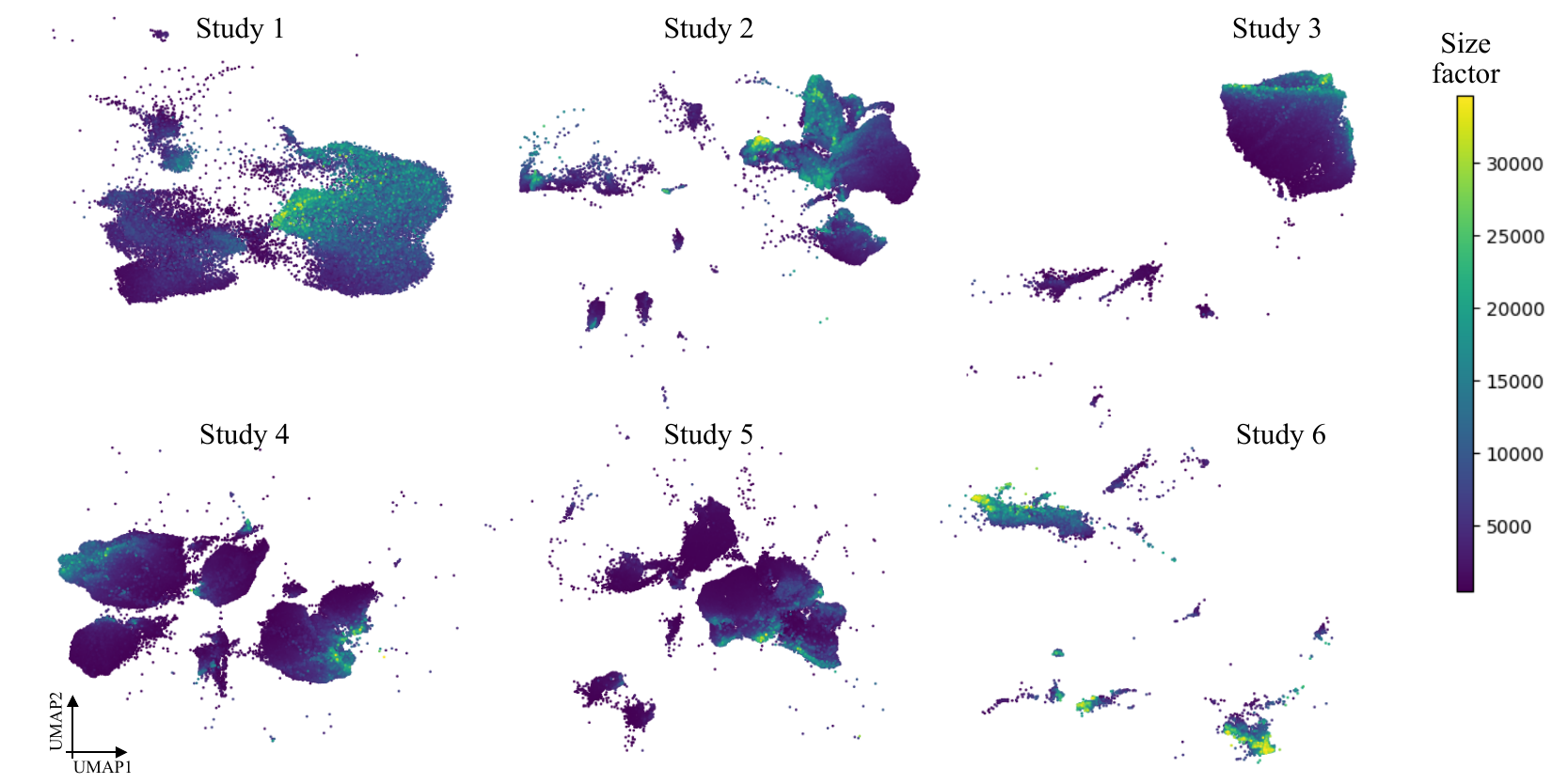}
\caption{UMAP plots of six studies included in the HLCA dataset colored by size factor.}
\label{fig: size_f_hlca}
\end{figure}

If the size factor is relatively uniform within a target population of cells, the approach in scVI might be sufficient. However, if we generate conditioned on a coarse annotation (e.g., the source study of a dataset in modern atlases), the library size can vary significantly within the annotation category (see \cref{fig: size_f_hlca} for examples using different studies in the Human Cell Atlas dataset). Sampling a latent state and size factor independently in such cases may lead to inconsistencies—scaling a decoded cell state by an incompatible size factor could produce unrealistic results.

Instead, conditioning on the size factor is more appropriate, as it biases the sampling of the latent state toward regions of the latent space where that specific size factor is naturally represented. This strategy can be combined with coarsely annotated variables to enable more targeted conditional generation.

\section{Model details}\label{sec: model_details}
The CFGen model is implemented in \texttt{PyTorch} \citep{paszke2017automatic}, version \texttt{2.1.2}.

\subsection{The CFGen Autoencoder}\label{sec: encoder_model}
Before training the flow model to generate noise from data, we first embed the data using an autoencoder trained with maximum likelihood optimization. 

\paragraph{Encoder.} The encoder is a multi-layer perceptron (MLP) with two hidden layers of dimensions \texttt{[512, 256]}. The final layer maps the input to a latent space, whose dimensionality is dataset-dependent. In our experiments, we use a 50-dimensional latent space for most datasets, except for the Human Lung Cell Atlas (HLCA) and Tabula Muris, where we set the latent space to 100 dimensions for increased representational capacity. In the multi-modal setting, different data modalities are embedded into a shared latent space. Each modality is first processed by a modality-specific MLP encoder. Due to the high dimensionality of DNA accessibility data (referred to as ATAC data from Assay for Transposase-accessible Chromatin), its encoder uses hidden layers of dimensions \texttt{[1024, 512]}. The outputs of the RNA and ATAC encoders are then concatenated and passed through a shared encoder layer, mapping to a final 100-dimensional latent representation.

\paragraph{Decoder.} The decoder maps the latent space to the parameter space of a likelihood model. For multi-modal data, each modality has its dedicated decoder.

\begin{itemize}
    \item scRNA-seq: The latent representation is mapped to the mean parameter $\bm{\mu}$ of a negative binomial likelihood, with one dimension per gene. Following \citet{Lopez2018}, we apply a \texttt{softmax} transformation across genes to produce normalized probabilities. These probabilities are then scaled by the library size (total transcript count per cell). The inverse dispersion parameter of the negative binomial distribution is a learned model parameter, implemented via \texttt{torch.nn.Parameter}. We offer the option to model inverse dispersion per gene or gene-attribute pair, depending on dataset properties.
    
    \item ATAC-seq: The decoder maps the latent space to continuous logit values. These are passed through an elementwise \texttt{sigmoid} function to produce probabilities per genomic region. Unlike RNA data, no size factor scaling is required.
\end{itemize}

\paragraph{Additional training details.}
We train the encoder and decoder networks jointly via likelihood optimization. Therefore, the weights of the networks are optimized (along with the inverse dispersion parameter) to produce the parameters that maximize the likelihood of the data under a predefined noise model. For scRNA-seq, we employ a negative binomial distribution, while for ATAC-seq, we use a Bernoulli likelihood. The losses from different modalities are summed before applying backpropagation. Note that scRNA-seq data are provided by the encoder in their $\log$-transformed version for training stability. However, the loss is evaluated on the original count data.

For all settings, we set the learning rate to $0.001$, with all layer pairs interleaved with one-dimensional batch normalization layers. We use the \texttt{AdamW} optimizer and the \texttt{ELU} activation function as the non-linearity.

\subsection{The velocity model}\label{sec: flow_model}

\paragraph{The vector field architecture.}
The vector field model takes as input the \textit{latent representation} computed by the encoder and produces a vector field used to simulate paths that generate data from noise. The architecture follows a deep, ResNet \citep{he2016deep} whose output has the same number of features as the input. The velocity model consists of the following components:
\begin{itemize}
    \item A linear projection layer that maps the input dimension to the hidden dimension of the flow model.
    \item Three stacked ResNet blocks are responsible for representation learning and conditioning. 
    \item An output layer with a single non-linearity, implemented using a \texttt{SiLU} activation function. 
    \item A time embedder that encodes time using sinusoidal multi-dimensional embeddings \citep{NIPS2017_3f5ee243} with a frequency value of \texttt{1e4}. This embedder is an MLP with two layers and \texttt{SiLU} non-linearity.
    \item A size factor embedder that applies sinusoidal embeddings \citep{NIPS2017_3f5ee243} to guide generation towards a predefined number of transcripts. This component is used only when the size factor is a conditioning variable, i.e., when we do not assume $p(\z|y, l) = p(\z|y)$. Before being passed through the sinusoidal embeddings, the $\log$ size factor is normalized to a range approximately between 0 and 1, using the maximum and minimum $\log$ size factors in the dataset. 
    \item Covariate embeddings for all different conditioning attributes.
\end{itemize}

\paragraph{Additional technical details.}
During training, the covariate embeddings are summed elementwise with the time embedding and, if applicable, the size factor embedding. Thus, all embeddings are either designed to have the same dimensionality or are transformed to a common size. This summed representation is then passed as a single vector to the ResNet blocks. 

Both the summed conditioning embedding and the down-projected input are provided to the ResNet blocks, which consist of:
\begin{itemize}
    \item A non-linear input transformation of the state embedding. 
    \item A linear encoder for the covariate embedding.
    \item A non-linear output transformation. 
    \item A skip connection.
\end{itemize}

The outputs of the non-linear input transformation and the covariate encoder are summed and passed through the output transformation. The result is then added to the input of the ResNet via the skip connection, following the traditional residual block structure \citep{he2016deep}. All non-linear transformations are implemented as simple \texttt{[SiLU, Linear]} stacks.

In the standard setting, we train the flow model for $1,000$ epochs using the \texttt{AdamW} optimizer, a learning rate of $0.001$, and a batch size of $256$.

\subsection{Covariate embeddings}
Covariate embeddings are trainable \texttt{torch.nn.Embdding} layers of pre-defined size. In our experiments, we use size $100$ in most of the settings.

\subsection{Sampling from noise}\label{sec: sampling_from_noise}
To generate discrete observations from noise, we first draw a covariate from the associated categorical distribution, with proportions estimated from the observed data. Next, we sample a size factor from a $\mathrm{LogNormal}$ distribution, where the mean and standard deviation are set as the Maximum Likelihood Estimates (MLE) from the entire dataset or conditioned on a technical effect covariate. We then sample Gaussian noise and simulate a latent observation from the real dataset conditionally. This is done by integrating the vector field computed by the neural network in \cref{sec: flow_model}, starting from Gaussian noise. The integration is performed using the \texttt{dopri5} solver with \texttt{adjoint} sensitivity and a tolerance of \texttt{1e-5} from the \texttt{torchdyn} package \citep{politorchdyn} in Python3 \citep{py3}, over the $[0,1]$ time interval. The generated latent vector is then decoded into the parameter space of the noise model for the data—negative binomial for scRNA-seq or Bernoulli for ATAC-seq—after which single cells are sampled.

\subsection{Separate training}\label{sec: separate_training}
In CFGen, we train the encoder $f_{\eta}$ separately from the flow model. Initially, when we attempted to model the autoencoder and the flow jointly, we found that training the flow was unstable. Specifically, Flow Matching performs better when the state space is fixed. Alternating between autoencoder and flow updates leads to continuous changes in the data representation, as the autoencoder evolves with the flow. This dynamic hinders accurate velocity field estimation, particularly during the early updates of the autoencoder. One could initially train the autoencoder with a higher learning rate than the flow, periodically decreasing the former and increasing the latter. However, this approach is essentially similar to training the autoencoder and flow separately, which is the strategy we ultimately adopt to avoid the need for repeatedly retraining the autoencoder.

\subsection{Scheduling}
We use linear scheduling, following the original formulation from \citet{lipman2022flow}.

\section{Baseline description}\label{sec: baseline_descr}

\subsection{scVI, MultiVI, PeakVI}
scVI \citep{Lopez2018}, MultiVI \citep{ashuach2023multivi}, and PeakVI \citep{ashuach2022peakvi} are all VAE-based generative models designed for single-cell discrete data. Following the standard VAE framework, these models learn a Gaussian latent space, which is then decoded into the parameters of discrete likelihood models that describe different single-cell modalities. While scVI and PeakVI generate single modalities—scVI for scRNA-seq and PeakVI for ATAC data—MultiVI learns a shared latent space across modalities while maintaining separate discrete decoders for each data type.

\subsection{scANVI and scPoli}\label{sec: scanvi_and_scpoli}

In the batch correction experiment described in \cref{sec: batch_correction}, we compare CFGen with two additional VAE-based models: scANVI \citep{Xu2021} and scPoli \citep{DeDonno2023}. scANVI extends scVI by incorporating a latent cell type classifier to preserve biological structure in the representation space and introducing a conditional prior on the latent space. scPoli differs from scANVI by using continuous embeddings instead of one-hot encodings as conditioning inputs to the VAE. Additionally, it enforces biological coherence by aligning cellular representations with latent cell-type prototypes. In simple terms, scPoli encourages cells to cluster around the average embedding vector of their respective cell types.

\subsection{scGAN}

The scGAN model \citep{marouf2020realistic} is a Generative Adversarial Network (GAN) \citep{goodfellow2014generative} designed for realistic scRNA-seq data generation. It minimizes the Wasserstein distance between real and generated cell distributions, employing a generator network to produce synthetic samples and a critic network to distinguish real from generated cells. The architecture includes fully connected layers and a custom library-size normalization (LSN) layer for stable training. The model extends to conditional scGAN (cscGAN) for type-specific cell generation. Evaluation relies on metrics such as t-SNE visualization and marker gene correlation to assess the quality of generated cells. While scGAN has been explored for conditional generation, we found that conditioning on cell type led to significantly worse results compared to an alternative approach in which the model is conditioned on data-driven Leiden cluster labels rather than real cell-type labels. We refer to this alternative as the \textit{unconditional} version, as it does not use predefined labels but instead leverages cluster-derived attributes.

\subsection{scDiffusion}

scDiffusion \citep{luo2024scdiffusion} is a generative model that leverages diffusion models to generate realistic single-cell gene expression data. The model consists of three main components:  

\begin{itemize}
    \item \textbf{An autoencoder}, which maps gene expression profiles to a latent space, enabling compression and feature extraction from high-dimensional single-cell data.  
    \item \textbf{A diffusion backbone network}, which learns to reverse a diffusion process applied to the latent embeddings, progressively refining noisy representations into meaningful biological signals.  
    \item \textbf{A conditional classifier}, which guides the generative process by incorporating cell type or other biological attributes, ensuring controlled cell generation.  
\end{itemize}

During training, the autoencoder first encodes real single-cell data into a latent representation. Noise is then progressively added to these embeddings following a predefined diffusion schedule. The diffusion backbone network is trained to learn the reverse process, reconstructing clean embeddings from noisy ones. Simultaneously, the conditional classifier is optimized to predict labels from these latent representations, reinforcing biological relevance in the learned distribution.

At inference, scDiffusion starts from a random noise vector in the latent space and iteratively removes noise using the trained diffusion backbone, ultimately generating a clean embedding. This embedding is then decoded by the autoencoder to reconstruct synthetic gene expression data. The process enables controlled single-cell generation by conditioning on specific biological attributes.  

\subsection{Discussion: Key Differences Between scDiffusion and CFGen}  

\paragraph{Handling Single-Cell Data Properties.}  
scDiffusion applies Gaussian diffusion to preprocessed single-cell data, disregarding key properties such as sparsity, overdispersion, and discreteness. While normalization ensures data continuity, most single-cell methods preserve zeros and non-linear mean-variance trends. Since continuous decoders like scDiffusion require centered, dense inputs, their design is suboptimal for scRNA-seq data.  

\paragraph{Conditional Sampling.}  
scDiffusion relies on classifier-based guidance, making conditional generation highly dependent on classifier accuracy. This limits its ability to generate rare cell types or handle attributes that are difficult to classify.  

\paragraph{Efficiency and Sampling Speed.}  
CFGen is two to three orders of magnitude faster than scDiffusion due to:  
\begin{itemize}
    \item Efficient Sampling: Flow Matching directly maps noise to data along nearly straight paths, requiring only 5–10 integration steps compared to the >1000 steps needed for scDiffusion.  
    \item Lower Dimensionality: CFGen operates in a compact latent space (50–100 dimensions), whereas scDiffusion performs denoising in a much higher-dimensional space (1000 dimensions).  
    \item Guidance approach: Unlike scDiffusion, which relies on a classifier’s gradient, CFGen uses CFG-based guidance, avoiding performance bottlenecks due to classifier accuracy.  
\end{itemize}

\section{Data preprocessing and description}\label{preproc}
Single-cell data were preprocessed using \texttt{scanpy} \citep{wolf2018scanpy}. Count normalization was applied only to baseline models requiring real-valued inputs. In these cases, gene counts were library-size normalized to \texttt{1e4} and $\log$-transformed. Since CFGen, MultiVI, PeakVI, and scVI operate in discrete space, we trained them on raw counts without normalization. Additionally, we filtered out genes expressed in fewer than 20 cells across all datasets.

\begin{table}[]
\centering
\caption{List of datasets considered in this work with the associated number of genes, cells and cell types. }
\resizebox{0.7\textwidth}{!}{%
\label{tab: dataset_specifics}
\begin{tabular}{c|c|c|c}
\hline
Dataset name  & Number of cells & Number of genes & Number of cell types \\ \hline
PMBC3K        & 2,638           & 8,573           & 8                    \\ \hline
Dentate gyrus & 18,213          & 17,002          & 14                   \\ \hline
Tabula Muris  & 245,389         & 19,734          & 123                  \\ \hline
HLCA          & 584,944         & 27,997          & 50                   \\ \hline
PBMC10k       & 10,025          & 25,604          & 14                   \\ \hline
NeurIPS       & 90,261          & 14,087          & 45                   \\ \hline
PBMC COVID    & 422,220         & 2,000           & 29                   \\ \hline
C.Elegans    & 89,701         & 17,747           & 35 (plus unknown)                   \\ \hline
\end{tabular}}
\end{table}

\section{Experiment description and evaluation metrics}\label{sec: metrics_and_experiments}
\subsection{2-Wasserstein distance and MMD}
We use the 2-Wasserstein distance and the RBF-kernel Mean Maximum Discrepancy (MMD) with scales $\{0.01, 0.1, 1, 10, 100\}$ \citep{gretton_mmd} to measure the overlap between real and generated data. To implement the former we use the Python Optimal Transport (POT) \citep{flamary2021pot} package. For the MMD, we resort to the implementation proposed in \footnote{\url{https://github.com/atong01/conditional-flow-matching}}.

\subsection{Distribution Metrics Comparisons}\label{sec: dist_mat_comp_app}  
To compute the metrics in \cref{tab:performance_comparison}, we generate three datasets per model, each matching the size of the original. In the conditional setting, we compute distribution metrics per cell type, comparing subsets of real and generated data. In the unconditional setting, we sample batches of $5,000$ cells from the full distribution. To mitigate the curse of dimensionality, we compute MMD and 2-Wasserstein distances in a 30-dimensional Principal Component (PC) space. Generated cells are projected using PC loadings from real cells to ensure comparability. Since scDiffusion and scGAN generate normalized data while CFGen and scVI produce discrete counts, we normalize CFGen and scVI outputs to a total of \texttt{1e4} counts per cell and then apply a $\log$-transformation. The same processing is applied to real data, ensuring all models operate on comparable quantities. All metrics are reported on the test set.  

\subsection{Variance-Mean Trend and Sparsity Histograms}  
After pre-processing (as described in \cref{sec: dist_mat_comp_app}), we compute the mean and variance of gene expression across cells and the frequency of unexpressed genes per cell. The mean-variance trend in raw count data is expected to be quadratic (see \cref{sec: gamma_poisson_negative_binomial}). However, normalization and $\log$-transformation—required for comparison with scDiffusion and scGAN—alter this trend. Despite this, examining the empirical mean-variance relationship remains informative, as it should align with real data behavior.

\subsection{Multi-Modal Evaluation}\label{sec:multimodal_eval}  
We generate multi-modal data and perform an unconditional comparison with the ground truth, as described in \cref{sec: dist_mat_comp_app}. For ATAC data, we normalize both real and generated cells using the TF-IDF algorithm from the MUON package \citep{bredikhin2022muon}.  

\begin{figure}[H]
\centering
\includegraphics[width=1\textwidth]{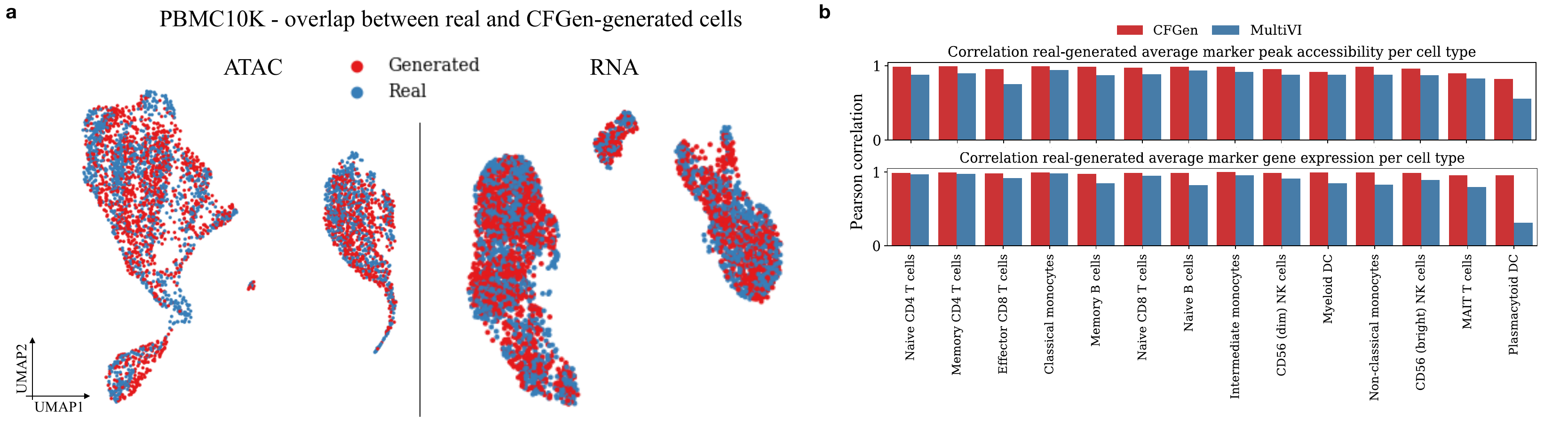}
\caption{\textbf{(a)} 2D UMAP overlap between real and generated cells across modalities on the PBMC10K dataset. \textbf{(b)} Pearson correlation between average cell-type-specific marker peak accessibility and marker gene expression between real data and samples from CFGen and MultiVI.}
\label{fig: multimodal_figure}
\end{figure}

To generate \cref{fig: multimodal_figure}b, we follow these steps:  
\begin{enumerate}  
    \item Compute the average gene expression and peak accessibility (i.e., fraction of accessible regions) per cell type for marker genes/peaks, following \footnote{\url{https://muon-tutorials.readthedocs.io/en/latest/single-cell-rna-atac/pbmc10k/3-Multimodal-Omics-Data-Integration.html}}. This yields a \texttt{cell\_type x marker} matrix for both real and generated datasets, where each entry represents the mean expression or accessibility of a marker in a given cell type.  
    \item Correlate the row vectors of these matrices between real and generated datasets. A high correlation indicates that the generative model accurately captures mean marker expression and accessibility per cell type.  
\end{enumerate}

\subsection{Guidance strength experiments}
\cref{fig: multi_label_guidance} illustrates the qualitative performance of guidance in CFGen. First, we train CFGen on each dataset following \cref{alg:guided_flows_tr}. Upon successful training, we sample 500 cells under different guidance strength combinations, varying one attribute while keeping the other fixed, as shown in the figure. For unconditional generation (i.e., guidance strength of 0 for both attributes), we generate as many cells as in the real dataset to better visualize the overlap between real and synthetic data. When applying guidance, we train the guided CFGen model with an unconditional sampling probability of $p_{\mathrm{uncond}}=0.2$ (see \cref{alg:guided_flows_tr}).

\subsection{scGPT \citep{cui2024scgpt} generalization performance enhancement}  
We split the PBMC COVID and HLCA datasets into a training set and a held-out set, ensuring a more challenging generalization task by leaving out all cells from 20\% of the donors in both datasets. This results in 80 training and 27 test donors for HLCA and 60 training and 15 test donors for PBMC COVID. After augmenting the training set (see \cref{sec: results - augmentation}), we use a pre-trained scGPT model to embed both the training and validation sets for both the original and augmented data. We then fit a k-Nearest-Neighbor (kNN) classifier on the training embeddings and evaluate its performance on the held-out set. The results, shown in \cref{fig: scgpt_exp}, illustrate how the classification performance in terms of recall on a cell type varies as a function of its frequency in the dataset after augmentation. An improvement in performance suggests that the additional synthetic examples help the model better distinguish the cell type in the LLM representation space.

\subsection{Batch Correction Evaluation}\label{sec: batch_correction_appendix}

\paragraph{Correction with CFGen.}  
Batch correction aims to remove technical effects while preserving biological variation in single-cell data. Given an observation $\mathbf{x}$ associated with a batch label $y_{\mathrm{batch}}$ and a biological annotation $y_{\mathrm{cell \, type}}$, we first encode $\mathbf{x}$ into a latent variable $\mathbf{z}$.  

Flow Matching is a generative model that maps a prior distribution to the data distribution via an invertible flow. This invertibility allows data distributions to be transported to noise. Inverting the flow back to the prior removes batch effects as well as cell-type variability from $\mathbf{z}$ \citep{rombach2020network}.  

To perform batch correction, we first apply flow inversion to remove both biological and technical variation from $\mathbf{z}$. Then, we simulate the flow forward again, starting from the noise representation. Given a reference batch $y^{\mathrm{ref}}_{\mathrm{batch}}$, we generate new observations conditioned on this batch while preserving the original biological label $y_{\mathrm{cell \, type}}$. When applied across the dataset, this procedure aligns all observations to the same batch, effectively removing batch-specific variations.  

\paragraph{Remarks.}  
In the context of CFG, the weights $\omega_{\mathrm{batch}}$ and $\omega_{\mathrm{cell \, type}}$ control the degree of biological preservation. Notably, our batch correction approach is conceptually similar to style transfer methods in diffusion models \citep{wang2023stylediffusion}.  

\paragraph{Evaluation Setup.}  
We train CFGen and all competing models using identical cell type and batch covariates. For a fair comparison, we use a latent space of 50 dimensions for all models. For uncorrected data, batch mixing is evaluated in the PC space.  

All VAE-based models were trained for 100 epochs with default settings, while scPoli was pre-trained for 40 steps, following \citet{DeDonno2023}.  

\paragraph{Metrics.}  
To assess batch correction and biological conservation, we use the scIB package \citep{Luecken2021}. Specifically, we employ five metrics for batch correction and five for biological conservation. The scores reported in \cref{tab: batch_correction_metrics} correspond to the average of these metrics, as computed by scIB. All scores are normalized between 0 and 1, where 1 indicates perfect correction or conservation.  

All metrics rely on kNN graphs, using batch and cell-type labels to evaluate technical and biological mixing. Below, we briefly describe each metric, though we refer to \citet{Luecken2021} for further details.  

\begin{enumerate}

\item \textbf{Batch Correction Metrics:}
\begin{itemize}
    \item Silhouette Batch – Measures the Average Silhouette Width (ASW) between batch clusters.
    \item iLISI – Computes the Inverse Simpson’s Index based on neighborhood composition in kNN graphs, indicating batch mixing quality.
    \item KBET – Evaluates whether the local batch composition in a cell’s kNN neighborhood matches the expected global batch distribution.
    \item Graph Connectivity – Assesses whether cells sharing the same label form a fully connected subgraph in the kNN representation.
    \item PCR (Principal Component Regression) – Quantifies batch-associated variance before and after correction.
\end{itemize}

\item \textbf{Biological Conservation Metrics:}
\begin{itemize}
    \item Isolated Labels – Identifies rare cell types appearing in the fewest number of batches and assesses their separation from other cell identities.
    \item K-means NMI – Computes the Normalized Mutual Information (NMI) between k-means clustering and batch clusters.
    \item K-means ARI – Measures the Adjusted Rand Index (ARI) between k-means clustering and batch clusters.
    \item Silhouette Label – Represents the Average Silhouette Width (ASW) between cell type clusters.
    \item cLISI – A cell-type-specific version of the iLISI score, evaluating biological structure preservation.
\end{itemize}

\end{enumerate}

\paragraph{Selection of the guidance weights for batch correction.} \cref{app:guidance_selection} provides an intuition for the selection process. In batch correction, cells are transported to noise and back to data guided by biological and batch covariates. The guidance strength parameters $\omega_{\mathrm{bio}}$ and $\omega_{\mathrm{batch}}$ determine the emphasis on biological conservation and batch correction. Based on the scIB metrics only, one might select the highest guidance strengths, as these maximize aggregation within cell types and batches. However, as shown in \cref{fig: corrected_a} and \cref{fig: corrected_b}, scIB metrics alone can be misleading and should be paired with qualitative evaluation. Excessive guidance collapses variability beyond the batch and biological annotations, leading to artefacts. Parameters near $\omega_{\mathrm{bio}}, \omega_{\mathrm{batch}} \in \{1, 2\}$ generally balance signal preservation and correction effectively. For example,  \cref{fig: corrected_b} demonstrates that excessive biological preservation causes unnatural clustering. The extent of batch effect in the data should also guide parameter selection. For C. Elegans, with mild batch effects, $\omega_{\mathrm{bio}}=2, \omega_{\mathrm{batch}}=1$ performs better than $\omega_{\mathrm{bio}}=1, \omega_{\mathrm{batch}}=2$. Conversely, for NeurIPS, $\omega_{\mathrm{bio}}=1, \omega_{\mathrm{batch}}=2$ avoids artifacts observed for $\omega_{\mathrm{bio}}>1$ (\cref{fig: corrected_b}). In summary, we recommend assessing the batch effect severity, sweeping over guidance weights, and selecting parameters that optimize scIB metrics without compromising realistic single-cell representations.

\section{Algorithms}\label{sec: algos}
\cref{alg:guided_flows_tr} and \cref{alg:sampling_guided_flows} depict our training strategies. In what follows, for notational simplicity, we indicate $\phi_t(\mathbf{z})$ with $\mathbf{z}_t$, where $t\in[0,1]$.

\begin{algorithm}[H]
\caption{Train CFGen with multiple attributes on scRNA-seq}
\label{alg:guided_flows_tr}
\begin{algorithmic}[1]
\REQUIRE Probability of unconditional generation $p_{\mathrm{uncond}}$, trained encoder $f_\eta$, scheduling $(\alpha_t, \sigma_t)$.
\STATE Initialize $v_{t,\xi}$ 
\WHILE {not converged}
    \STATE Sample $(\x, y_1,...,y_{K})$ from the data
    \STATE $\z_1 \leftarrow f_\eta(\x)$
    \STATE Sample $t$ from $\mathcal{U}[0,1]$
    \STATE $l \leftarrow \textrm{Sum of entries of }
    \x$
    \STATE Sample $b$ from $\mathrm{Bernoulli}(p_{\mathrm{uncond}})$
    \IF{$b$ = 1}
        \STATE $y \leftarrow \emptyset $
    \ELSE
        \STATE $y \leftarrow \textrm{sample uniformly a label among} \: y_1,...,y_{K}$
    \ENDIF    
    \STATE $\z_0 \sim \mathcal{N}(\mathbf{0},\mathbf{I})$ \COMMENT{sample noise}
    \STATE $\z_t \leftarrow \alpha_t \z_1 + \sigma_t \z_0$ \COMMENT{noisy data point}
    \STATE $\dot{\z}_t \leftarrow \dot{\alpha}_t \z_1 + \dot{\sigma}_t \z_0$ 
    \STATE Take gradient step on $\nabla_{\xi} || v_{t,\xi}(\z_t, y, l) - \dot{\z}_t||^2$
\ENDWHILE
\end{algorithmic}
\textbf{Output:} $v_{t,\xi}$
\end{algorithm}

\begin{algorithm}[H]
    \caption{Sampling from multi-attribute guided CFGen for scRNA-seq}
\label{alg:sampling_guided_flows}
\begin{algorithmic}[1]
\REQUIRE Trained velocity field $v_{t,\xi}$, conditions $y_1,...,y_{K}$, guidance parameters $\omega_1,...,\omega_{K}$, \\
size factor distribution parameters $(\mu_l, \sigma_l)$, number of ODE steps $n_{\textrm{ode}}$,\\ trained decoder $h_\psi$, trained inverse dispersion parameter $\bm{\theta}$.
\STATE Sample size factor $l$ from $\mathrm{LogNormal}(\mu_{l}, \sigma_{l})$
\STATE $\z_0 \sim \mathcal{N}(\mathbf{0},\mathbf{I})$ \COMMENT{sample noise}
\STATE $n \leftarrow 1/n_{\textrm{ode}}$ \COMMENT{step size}
\STATE $\tilde{u}_{t}(\cdot) \leftarrow v_{t,\xi}(\cdot, \emptyset, l) + \sum_{i=1}^{K} \omega_i \left[ v_{t,\xi}(\cdot, y_i, l) - v_{t,\xi}(\cdot, \emptyset,l) \right]$ \COMMENT{guided velocity function}
\FOR{$t = 0, n, \ldots, 1 - n$}
    \STATE $\z_{t+n} \leftarrow \textrm{ODEStep}(\tilde{u}_{t}, \z_t)$ \COMMENT{ODE solver step}
\ENDFOR
\STATE $\x \leftarrow \textrm{Sample from NB}(l \, \mathrm{softmax}(h_\psi(\z_1))
, \bm{\theta})$
\end{algorithmic}
\textbf{Output:} $\x$
\end{algorithm}

\clearpage
\section{Additional results}
\subsection{Analysis of the runtime}

\begin{figure}[H]
\centering
\includegraphics[width=0.65\textwidth]{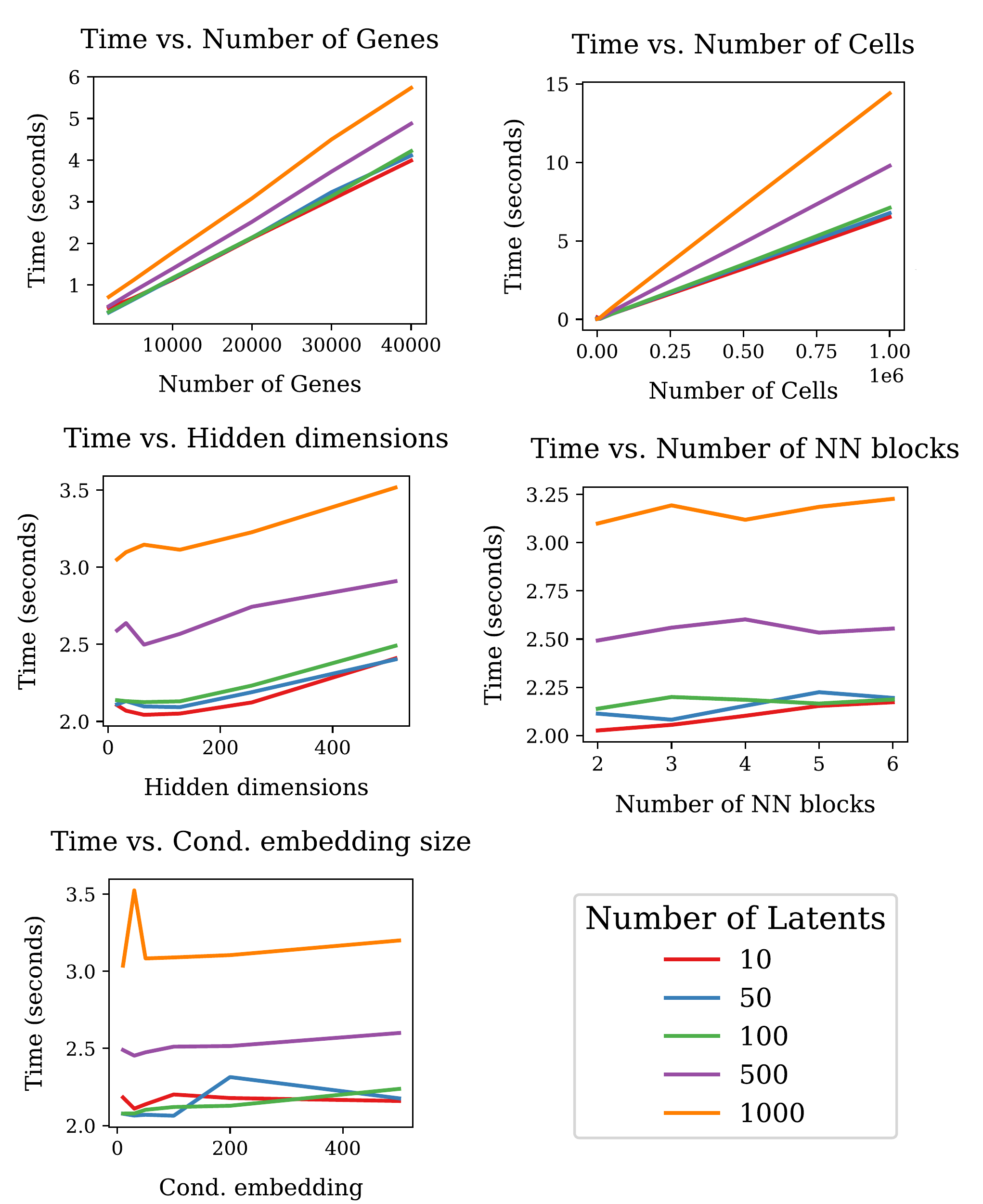}
\caption{Runtime analysis of the CFGen generation process. Each panel represents a different hyperparameter configuration. Different lines in each plot correspond to varying latent space dimensionalities, which directly impact generation time. We examine how runtime changes as a function of five key hyperparameters: (1) number of generated genes, (2) number of generated cells, (3) hidden dimension of the denoising model, (4) number of blocks in the denoising model and (5) size of the condition embedding. Results are reported in seconds. When varying one hyperparameter, all others are fixed at their default values.}
\label{fig: runtime_plot}
\end{figure}

In \cref{fig: runtime_plot}, we empirically evaluate how different hyperparameters impact CFGen’s runtime. To do so, we generate synthetic data using an untrained CFGen instance initialized with a specific configuration and run our experiments on an NVIDIA A100 GPU. Each hyperparameter is assessed across different latent space sizes, as the latent space dimension serves as a bottleneck in Flow Matching models and is expected to have the greatest influence on generation speed.

We consider the following hyperparameters:  
\begin{enumerate}
    \item Number of generated genes (default: 20,000).  
    \item Number of generated cells (default: 50,000).  
    \item Latent space dimensionality of the denoising model's bottleneck (default: 128).  
    \item Number of neural network blocks in the denoising model (default: 3).  
    \item Embedding size for conditional inputs (default: 128).  
\end{enumerate}  

When evaluating the effect of a specific hyperparameter, all others are fixed at their default values to ensure a controlled comparison.

In \cref{fig: runtime_plot}, we observe that the most influential hyperparameters affecting generation speed are the number of cells and genes, while factors related to the neural network size have a smaller impact. As expected, the number of latent codes significantly influences the sampling speed, as it determines the dimensionality of the generation space. Additionally, we compare the training and sampling runtimes of CFGen against competing models across different datasets (see \cref{tab:train_runtime} and \cref{tab:samp_runtime}).

\begin{table}[H]
\centering
\caption{Training runtime table. Each entry corresponds to the time in seconds required to train a model on different datasets. The number of cells and genes composing each dataset are reported at the bottom of the table. CFGen and scDiffusion are broken down into their different components that should be considered additively for an overview of the total runtime. For all the models, the batch size is set to 128.}
\label{tab:train_runtime}
\resizebox{0.9\textwidth}{!}{%
\begin{tabular}{cccccc}
\hline
                       & PBMC3K & Dentate gyrus & Tabula muris & HLCA    & PBMC10K (scRNA-seq) \\ \hline
CFGen Flow               & 1.02   & 7.13          & 69.00        & 192.12  & 3.23                \\
CFGen AE               & 1.40   & 6.31          & 68.40        & 253.21  & 6.30                \\
scVI                   & 0.08   & 2.11          & 18.13        & 65.12   & 2.15                \\
MultiVI                & -      & -             & -            & -       & 22.12               \\
scDiffusion Denoiser         & 1.03   & 2.13          & 18.02        & 53.62   & 4.48               \\
scDiffusion AE         & 0.98   & 7.14          & 165.6        & 329.02  & 7.39                \\
scDiffusion classifier & 0.01   & 0.71          & 9.66         & 26.32   & 0.39                \\
scGAN                  & 0.98   & 5.15          & 20.41        & 181.12  & 2.40                \\ \hline
\textbf{No. of cells}  & 2,638  & 18,213         & 245,389      & 584,944 & 10,025              \\
\textbf{No. of genes}  & 8,573  & 17,002         & 19,734       & 27,997  & 25,604              \\ \hline
\end{tabular}}
\end{table}

\begin{figure}[H]\label{comp}
\centering
\includegraphics[width=0.8\textwidth]{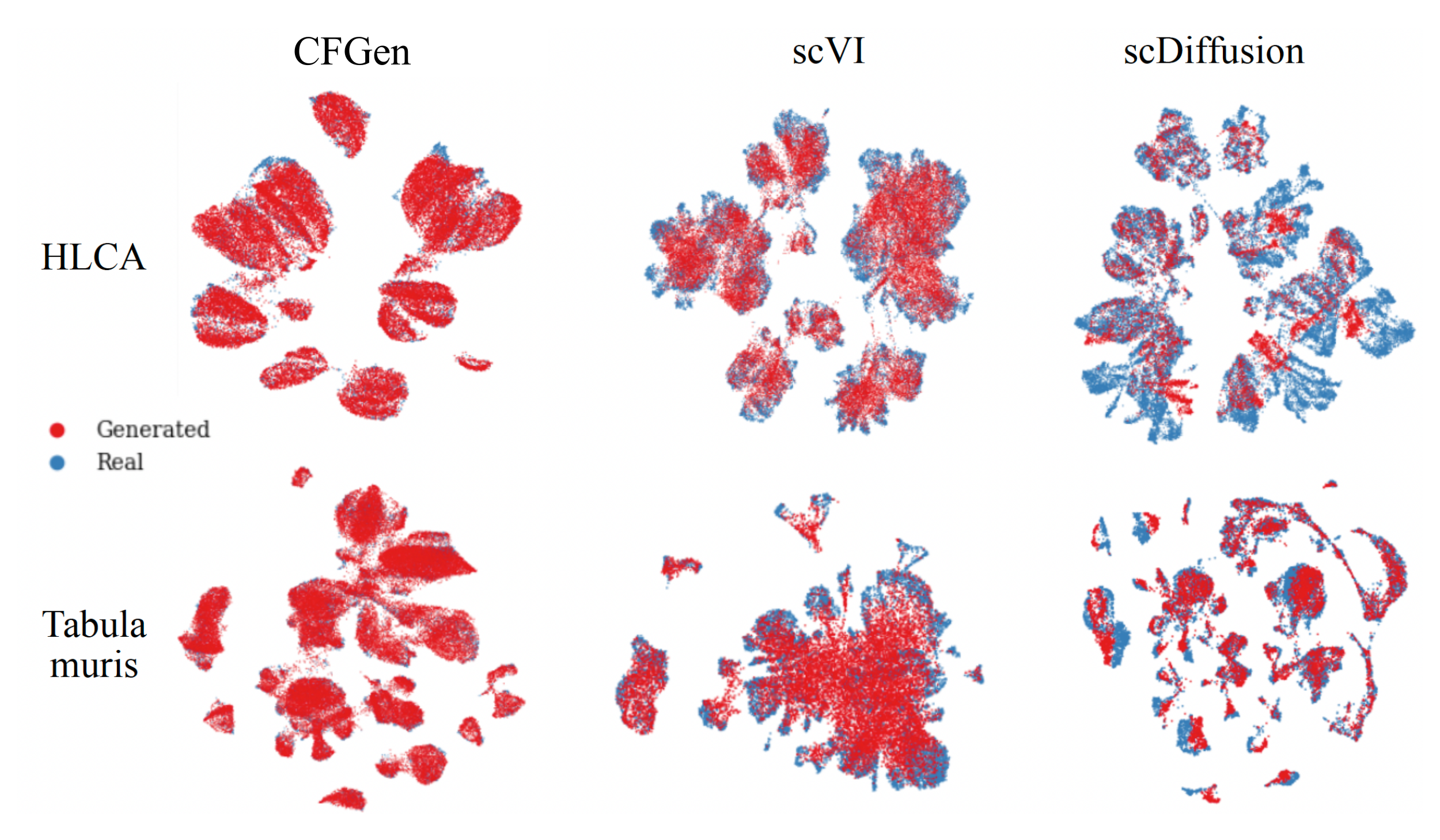}
\caption{Qualitative comparison of the generation results of CFGen, scVI and scDiffusion on the HLCA and Tabula Muris datasets. Comparison is performed by evaluating the similarity of the generated results to real cells. Real and generated cells for all models are embedded together and visualized as 2D UMAP coordinates.}
\label{fig: comp_gen}
\end{figure}

\begin{table}[H]
\centering
\caption{Generation runtime table. Each entry corresponds to the time in seconds required for a model to generate as many cells and genes as in the original dataset. The number of cells and genes composing each dataset are reported at the bottom of the table.}
\label{tab:samp_runtime}
\resizebox{0.8\textwidth}{!}{%
\begin{tabular}{cccccc}
\hline
                      & PBMC3K & Dentate gyrus & Tabula muris & HLCA    & PBMC10K (scRNA-seq) \\ \hline
CFGen              & 0.34   & 0.26          & 3.68         & 8.62    & 0.43                \\
scVI                  & 0.01   & 0.02          & 1.26         & 3.63    & 0.03                \\
MultiVI               & -      & -             & -            & -       & 0.03                \\
scDiffusion        & 48.79  & 105.08        & 1255.41      & 2004.00 & 113.41              \\
scGAN                 & 0.70   & 0.94          & 4.15         & 12.39   & 0.68                \\ \hline
\textbf{No. of cells} & 2,638  & 18,213         &      245,389         & 584,944 & 10,025              \\
\textbf{No. of genes} & 8,573  & 17,002         & 19,734       & 27,997  & 25,604              \\ \hline
\end{tabular}}
\end{table}
From the sampling runtime results in \cref{tab:samp_runtime}, we observe that VAE-based models (scVI and MultiVI) are generally faster. However, it is important to note that these models are inherently less expressive and perform worse than CFGen, particularly on large datasets (see \cref{tab:performance_comparison} and \cref{fig: comp_gen}). Notably, CFGen outperforms scDiffusion in speed, accelerating sampling by orders of magnitude. This improvement is attributed to the following factors:

\begin{itemize}
    \item CFGen requires fewer simulation steps than scDiffusion (5-10 steps in CFGen vs. >1000 in scDiffusion) while achieving superior empirical and quantitative results.
    \item CFGen operates in a lower-dimensional latent space (50-100 dimensions vs. 1000 dimensions for scDiffusion, as recommended in the manuscript).
    \item CFGen employs classifier-free guidance, whereas scDiffusion relies on classifier-based guidance. Consequently, CFGen's performance is not affected by the gradient of a classifier’s prediction at each step.
\end{itemize}

Strikingly, CFGen can generate comprehensive atlases with over 500,000 cells, such as HLCA, in just 8 seconds. For fairness, we acknowledge that the speedup depends on the batch size that can fit into memory during sampling (10k cells in our case).
\newpage

\subsection{Example of synthetic generation by CFGen}

\begin{figure}[H]
\centering
\includegraphics[width=0.9\textwidth]{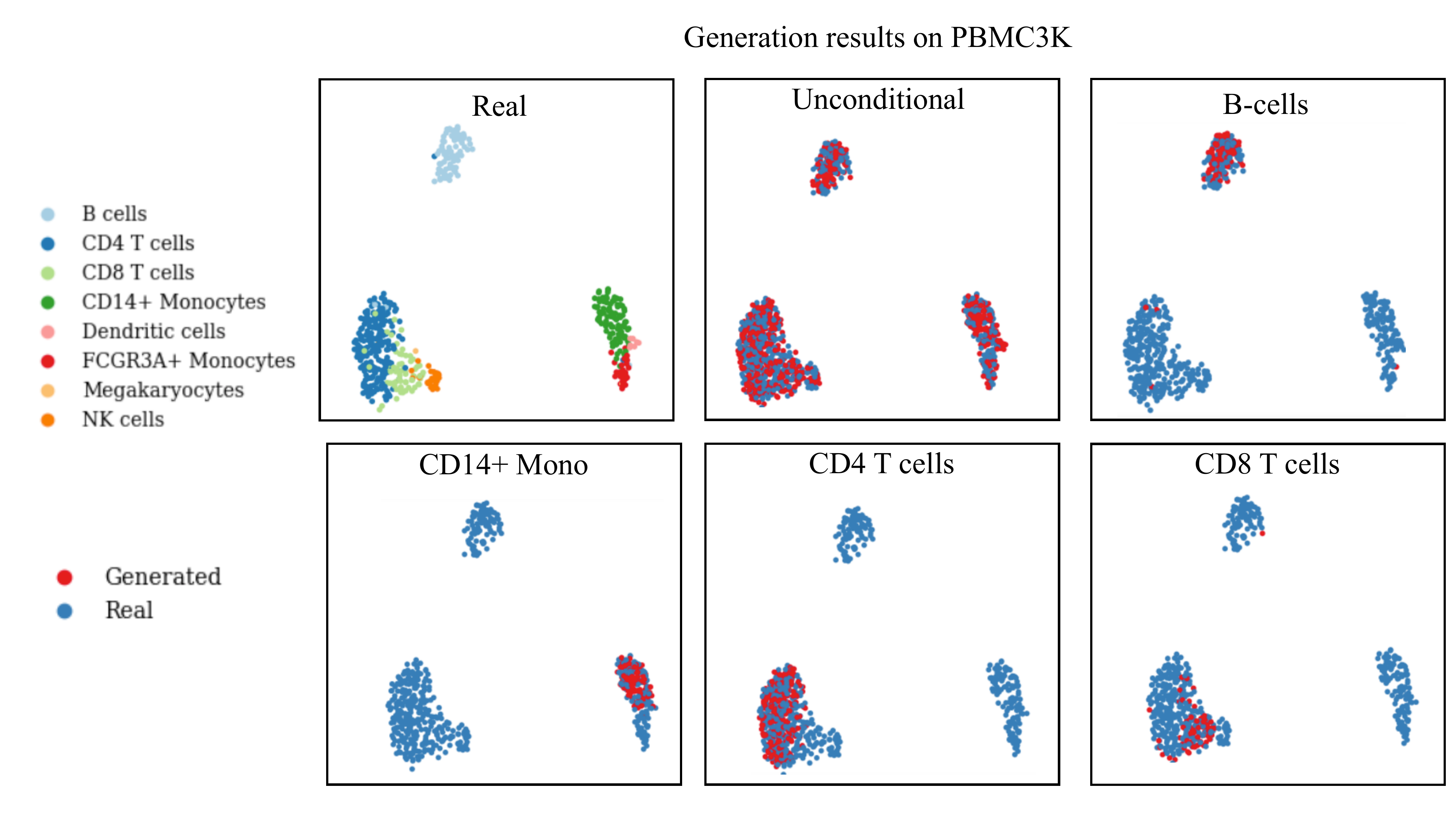}
\caption{Uni-modal generation of scRNA-seq by CFGen on the PBMC3K dataset. Real and generated cells are embedded together and visualized as 2D UMAP coordinates.}
\label{fig: pbmc_gen}
\end{figure}

\begin{figure}[H]
\centering
\includegraphics[width=0.9\textwidth]{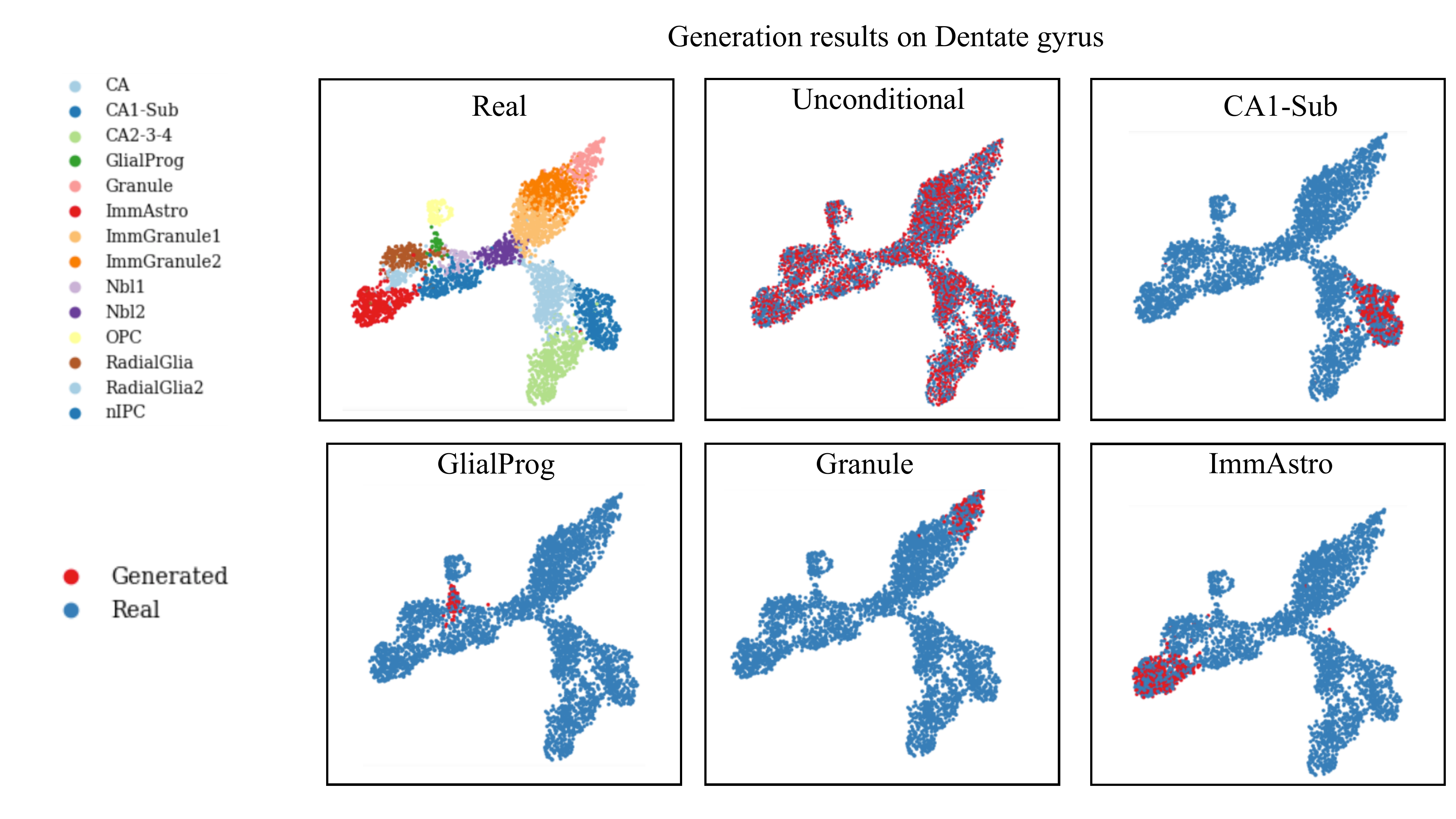}
\caption{Uni-modal generation of scRNA-seq by CFGen on the Dentate gyrus dataset. Real and generated cells are embedded together and visualized as 2D UMAP coordinates.}
\label{fig: dentate_gen}
\end{figure}
\newpage

\begin{figure}[H]
\centering
\includegraphics[width=0.9\textwidth]{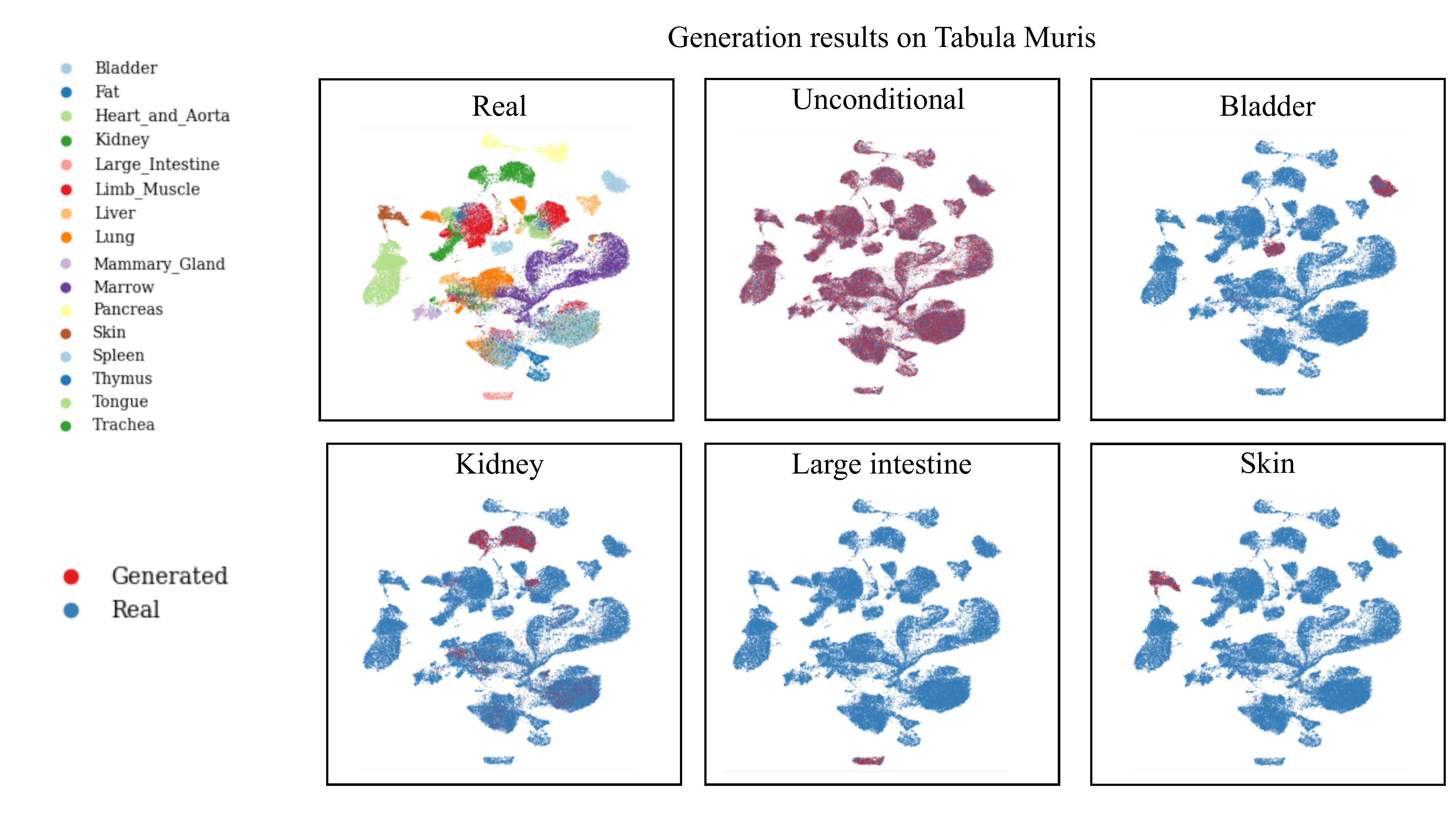}
\caption{Uni-modal generation of scRNA-seq by CFGen on the Tabula Muris dataset. Real and generated cells are embedded together and visualized as 2D UMAP coordinates.}
\label{fig: tabula_gen}
\end{figure}

\begin{figure}[H]
\centering
\includegraphics[width=1\textwidth]{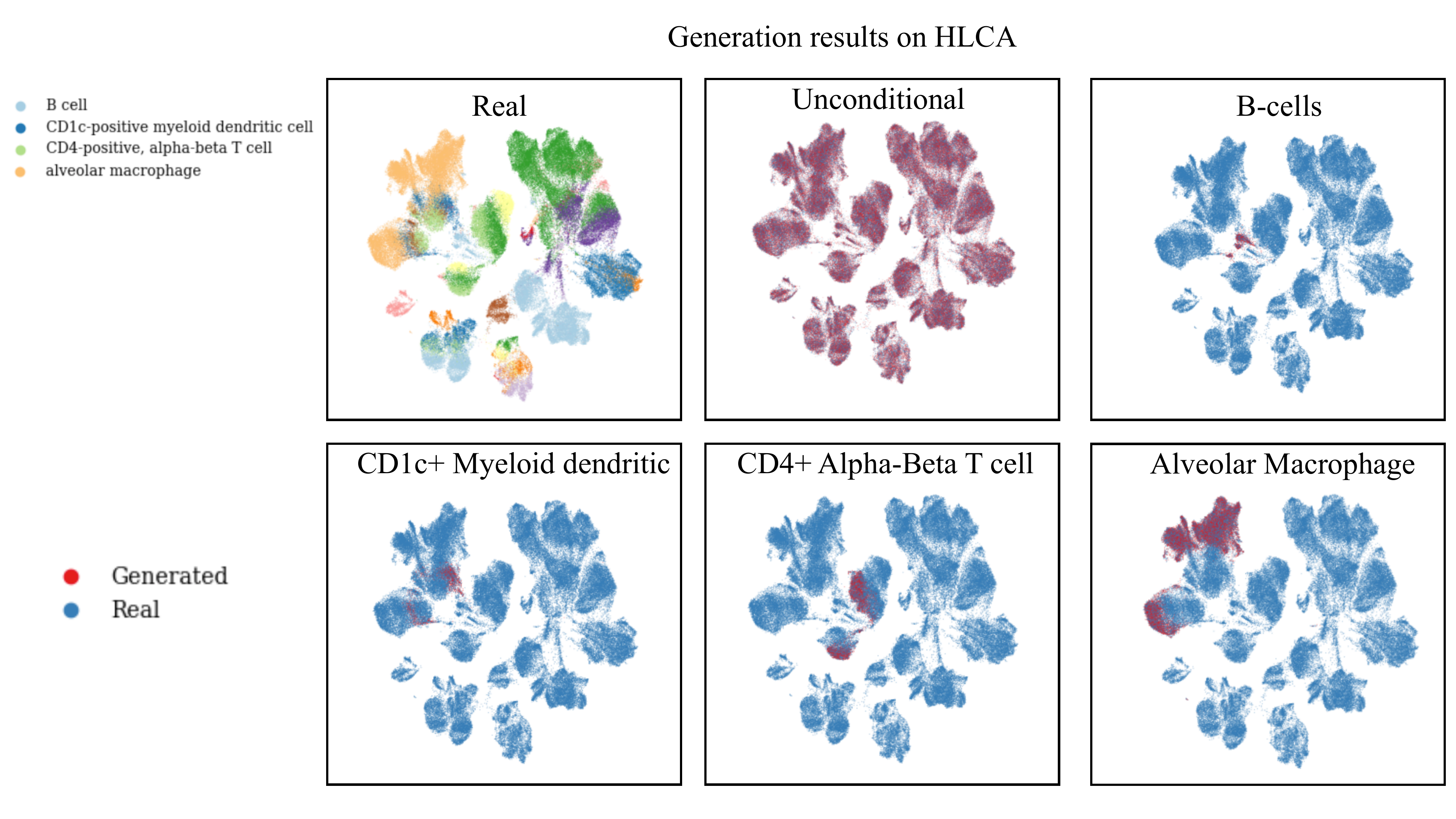}
\caption{Uni-modal generation of scRNA-seq by CFGen on the HLCA dataset. Real and generated cells are embedded together and visualized as 2D UMAP coordinates.}
\label{fig: hlca_gen}
\end{figure}
\newpage

\subsection{Comparison between CFGen, scVI and scDiffusion on the HLCA and Tabula Muris datasets}

\subsection{Additional results on multi-modal generation}\label{sec: additional_results_multimodal}

\newpage

\begin{figure}[H]
\centering
\includegraphics[width=0.85\textwidth]{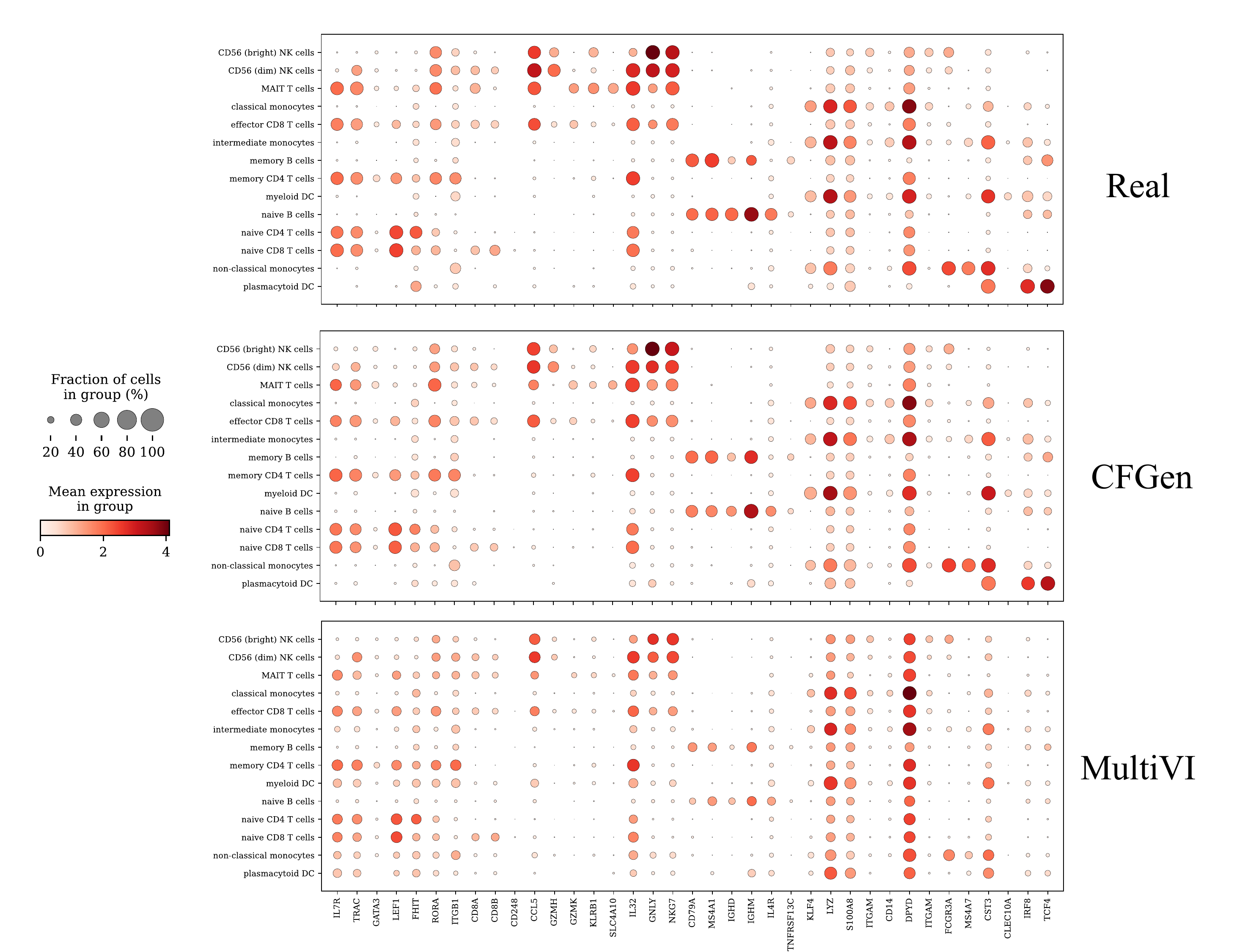}
\caption{Average marker expression per cell type in real and generated data on the PBMC10k dataset. \textbf{x-axis} - marker genes. \textbf{y-axis} - cell types.}
\label{fig: multimodal_figure_marker_genes}
\end{figure}

\begin{figure}[H]
\centering
\includegraphics[width=0.90\textwidth]{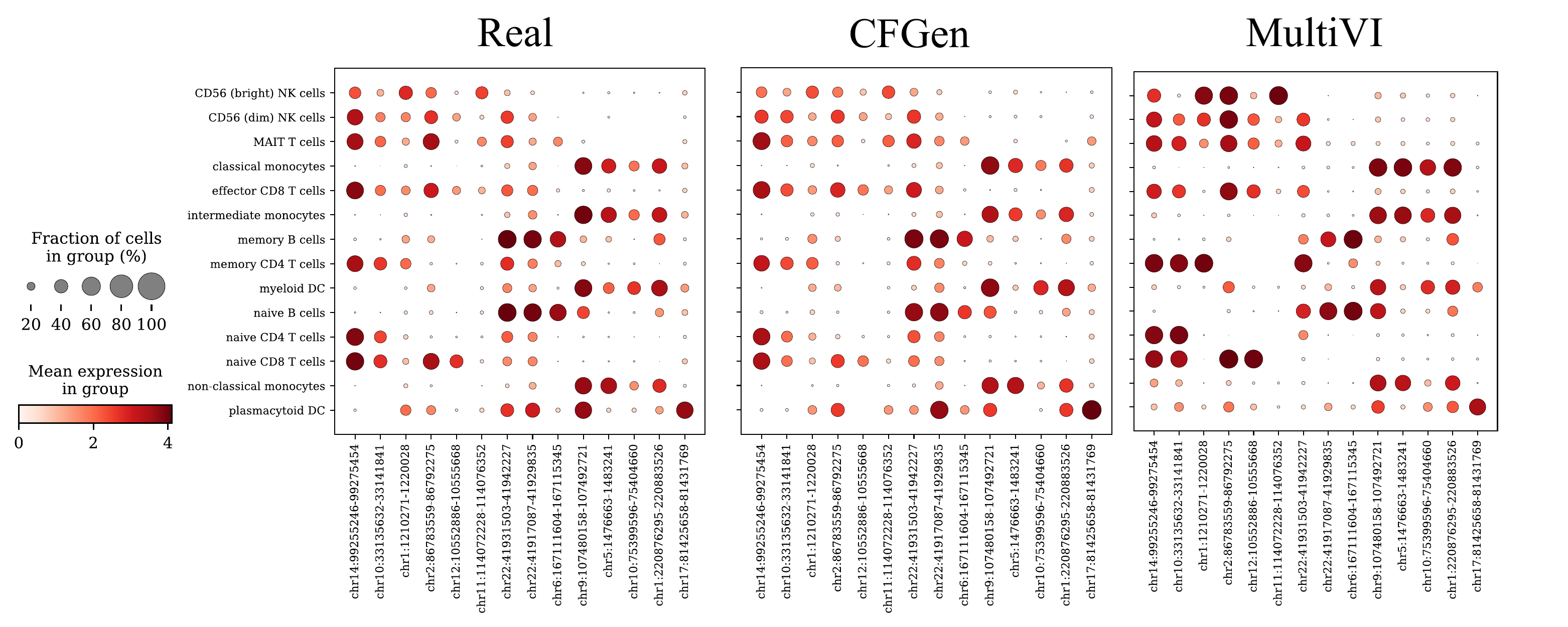}
\caption{Average number of cells with accessible marker peaks per cell type in real and generated data on the PBMC10k dataset. \textbf{x-axis} - marker peaks. \textbf{y-axis} - cell types.}
\label{fig: multimodal_figure_peaks}
\end{figure}

\newpage
\subsection{Additional results on data augmentation}

\begin{figure}[H]
\centering
\includegraphics[width=0.8\textwidth]{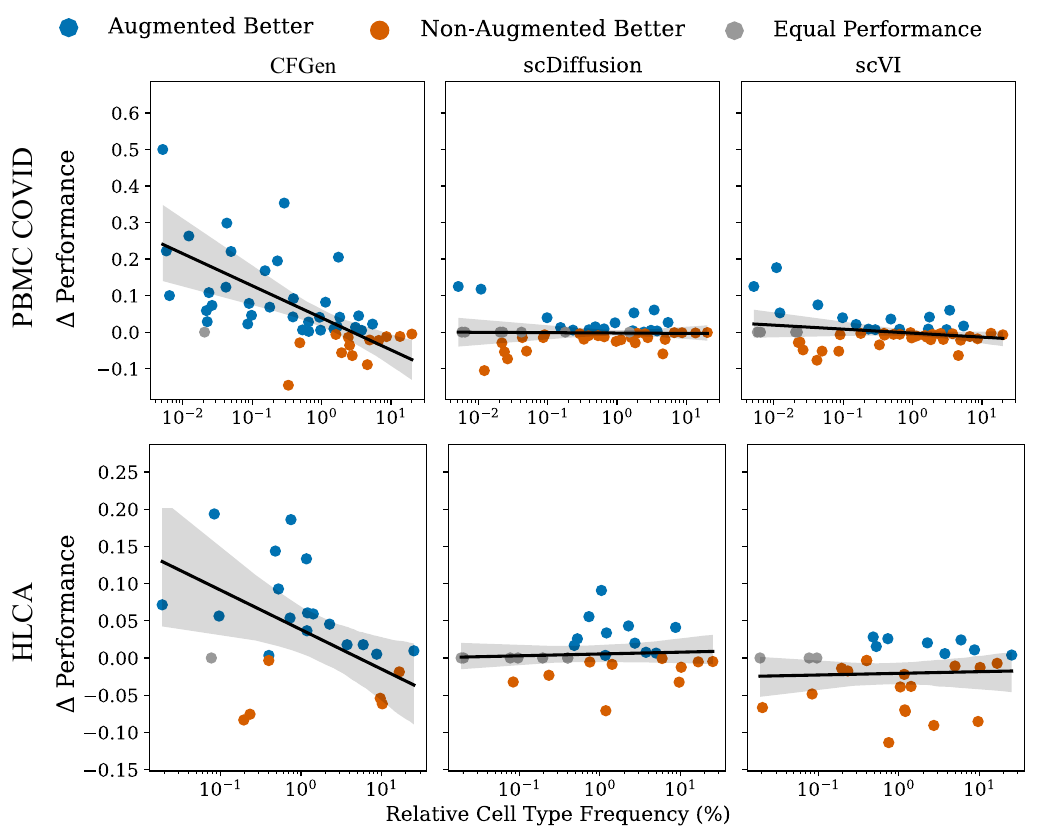}
\caption{Extension of \cref{fig: scgpt_exp}. Comparison of CFGen with scDiffusion and scVI on boosting the scGPT classifier performance in terms of recall on rare cell types.}
\label{fig: augmentation_comp}
\end{figure}

\begin{figure}[H]
\centering
\includegraphics[width=0.75\textwidth]{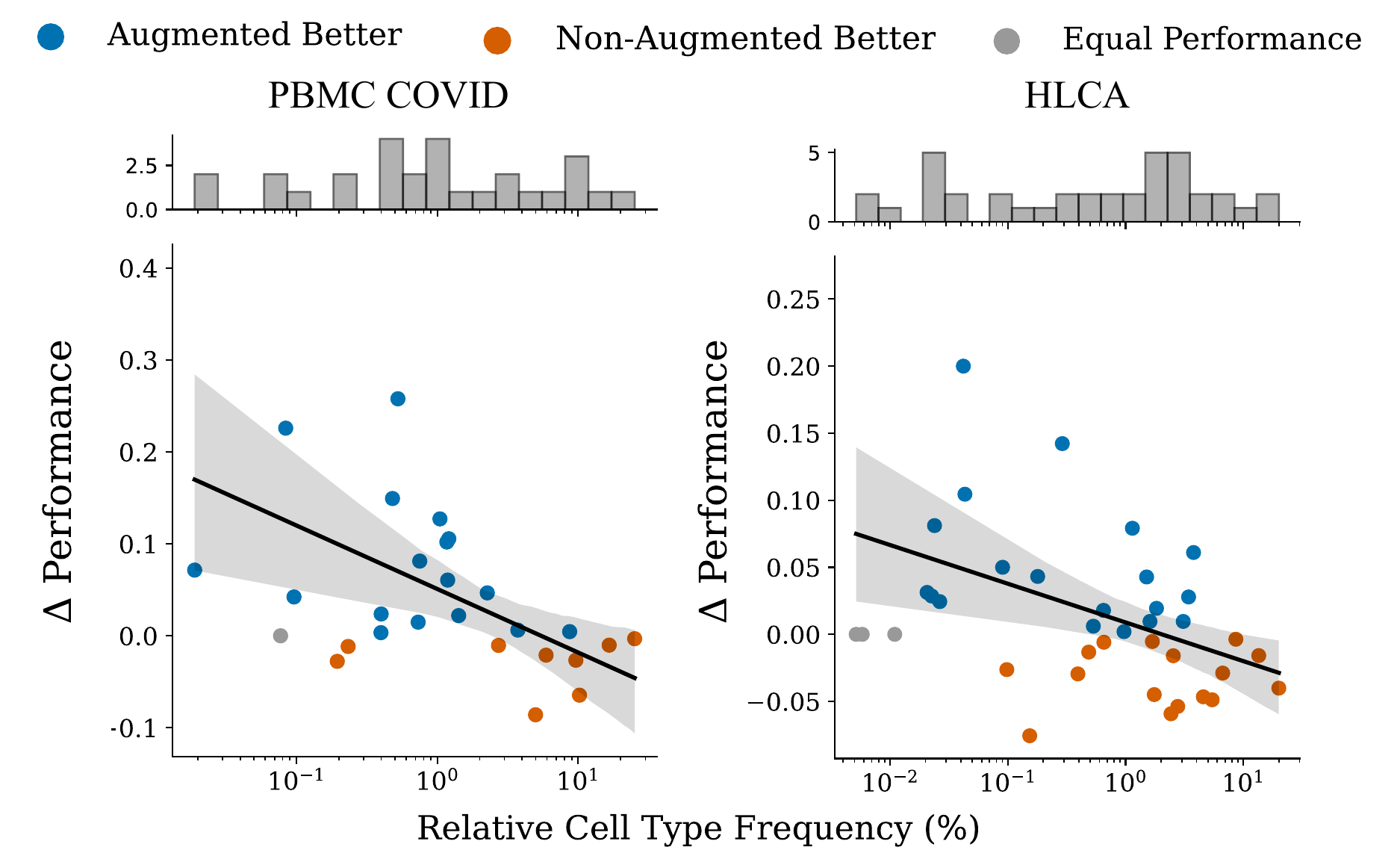}
\caption{Performance improvement of a linear cell type classifier after data augmentation. The plot displays the cell-type classification performance difference in terms of recall as a function of cell type frequency before and after augmentation on PBMC COVID and HLCA datasets. As a classifier, we use CellTypist \citep{cippa2023first}, which is based on logistic regression. The held-out set includes cells from 20\% of donors for both datasets.}
\label{fig: augmentation_comp_linear}
\end{figure}

Together with scGPT, in \cref{fig: augmentation_comp_linear} we investigate if using CFGen to augment individual cell types improves the performance of a linear classifier like CellTypist \citep{cippa2023first}. We obtain a similar result as scGPT, with the recall performance on real cell types improving after augmentation (hence a negative correlation between the performance improvement and the cell type frequency).

For a better appreciation of the classification improvement of single cell type categories by scGPT, we include \cref{tab:ind_acc_pbmc} and \cref{tab:ind_acc_hlca}.

\begin{table}[H]
\centering
\caption{Table reporting the cell type classification recall of scGPT before (Recall Base) and after (Recall Aug) augmentation on the PBMC covid dataset. The Relative Frequency (\%) column reports how rare a certain cell type is in the dataset. For each row, we highlight which setting leads to the highest recall.}
\label{tab:ind_acc_pbmc}
\resizebox{1\textwidth}{!}{%
\begin{tabular}{lccc}
\hline
\multicolumn{1}{c}{Cell Type}              & Relative Frequency (\%) & Recall Base & Recall Aug \\ \hline
naive thymus-derived CD4-positive, alpha-beta ...   & 25.18                            & 0.87                   & \textbf{0.90}         \\
classical monocyte                                  & 16.61                            & 0.98                   & 0.98                  \\
natural killer cell                                 & 10.22                            & \textbf{0.91}          & 0.90                  \\
CD4-positive helper T cell                          & 9.63                             & \textbf{0.80}          & 0.75                  \\
naive thymus-derived CD8-positive, alpha-beta ...   & 8.70                             & \textbf{0.83}          & 0.79                  \\
naive B cell                                        & 5.92                             & 0.96                   & \textbf{0.98}         \\
CD8-positive, alpha-beta cytotoxic T cell           & 4.98                             & \textbf{0.78}          & 0.67                  \\
non-classical monocyte                              & 3.73                             & 0.95                   & \textbf{0.96}         \\
central memory CD8-positive, alpha-beta T cell      & 2.72                             & \textbf{0.45}          & 0.33                  \\
regulatory T cell                                   & 2.27                             & 0.28                   & \textbf{0.32}         \\
conventional dendritic cell                         & 1.42                             & 0.79                   & \textbf{0.84}         \\
CD16-negative, CD56-bright natural killer cell ...  & 1.21                             & 0.66                   & \textbf{0.70}         \\
gamma-delta T cell                                  & 1.19                             & 0.49                   & \textbf{0.63}         \\
effector memory CD8-positive, alpha-beta T cell ... & 1.17                             & 0.21                   & \textbf{0.27}         \\
class switched memory B cell                        & 1.04                             & 0.51                   & \textbf{0.71}         \\
B cell                                              & 0.75                             & 0.21                   & \textbf{0.39}         \\
mucosal invariant T cell                            & 0.73                             & 0.69                   & \textbf{0.70}         \\
CD4-positive, alpha-beta cytotoxic T cell           & 0.53                             & 0.06                   & \textbf{0.13}         \\
effector memory CD8-positive, alpha-beta T cell     & 0.48                             & 0.10                   & \textbf{0.23}         \\
plasmacytoid dendritic cell                         & 0.40                             & 1.00                   & 1.00                  \\
platelet                                            & 0.40                             & 1.00                   & 1.00                  \\
plasma cell                                         & 0.23                             & \textbf{0.97}          & 0.90                  \\
hematopoietic precursor cell                        & 0.20                             & \textbf{0.97 }                  & 0.86         \\
mature NK T cell                                    & 0.10                             & 0.00                   & \textbf{0.16}         \\
innate lymphoid cell                                & 0.08                             & 0.21                   & \textbf{0.40}         \\
erythrocyte                                         & 0.08                             & 1.00                   & 1.00                  \\
dendritic cell                                      & 0.02                             & 0.47                   & \textbf{0.80}         \\
plasmablast                                         & 0.02                             & 0.93                   & \textbf{1.00}         \\
granulocyte                                         & 0.00                             & 1.00                   & 1.00                  \\ \hline
\end{tabular}}
\end{table}
\newpage

\begin{table}[H]
\centering
\caption{Table reporting the cell type classification recall of scGPT before (Recall Base) and after (Recall Aug) augmentation on the HLCA dataset. The Relative Frequency (\%) column reports how rare a certain cell type is in the dataset. For each row, we highlight which setting leads to the highest recall.}
\label{tab:ind_acc_hlca}
\resizebox{1\textwidth}{!}{%
\begin{tabular}{lccc}
\hline
Cell Type                              & Relative Frequency (\%) & Recall Base & Recall Aug \\ \hline
alveolar macrophage                             & 20.00                            & 0.95                   & 0.95                  \\
type II pneumocyte                              & 13.51                            & 0.99                   & 0.99                  \\
respiratory basal cell                          & 8.63                             & \textbf{0.92}                   & 0.90                  \\
ciliated columnar cell of tracheobronchial tree & 6.67                             & \textbf{0.97}          & 0.94                  \\
nasal mucosa goblet cell                        & 5.43                             & 0.88                   & \textbf{0.89}         \\
CD8-positive, alpha-beta T cell                 & 4.93                             & \textbf{0.89}          & 0.87                  \\
club cell                                       & 4.57                             & \textbf{0.62}          & 0.53                  \\
elicited macrophage                             & 3.77                             & 0.70                   & 0.70                  \\
CD4-positive, alpha-beta T cell                 & 3.43                             & 0.59                   & \textbf{0.64}         \\
vein endothelial cell                           & 3.09                             & 0.93                   & \textbf{0.94}         \\
capillary endothelial cell                      & 2.77                             & \textbf{0.92}          & 0.85                  \\
alveolar type 2 fibroblast cell                 & 2.54                             & \textbf{0.94}          & 0.90                  \\
classical monocyte                              & 2.43                             & 0.87                   & 0.87                  \\
CD1c-positive myeloid dendritic cell            & 1.95                             & \textbf{0.73}          & 0.64                  \\
pulmonary artery endothelial cell               & 1.83                             & 0.68                   & \textbf{0.75}         \\
lung macrophage                                 & 1.75                             & 0.36                   & \textbf{0.57}         \\
type I pneumocyte                               & 1.69                             & 0.94                   & \textbf{0.95}         \\
non-classical monocyte                          & 1.61                             & \textbf{0.55}          & 0.54                  \\
natural killer cell                             & 1.51                             & 0.81                   & \textbf{0.83}         \\
multi-ciliated epithelial cell                  & 1.14                             & 0.55                   & \textbf{0.66}         \\
endothelial cell of lymphatic vessel            & 0.97                             & 0.91                   & \textbf{0.93}         \\
epithelial cell of lower respiratory tract      & 0.94                             & 0.86                   & \textbf{0.88}         \\
mast cell                                       & 0.65                             & 0.96                   & 0.96                  \\
B cell                                          & 0.65                             & 0.88                   & \textbf{0.90}         \\
plasma cell                                     & 0.53                             & 0.98                   & 0.98                  \\
alveolar type 1 fibroblast cell                 & 0.49                             & \textbf{0.82}          & 0.79                  \\
bronchus fibroblast of lung                     & 0.39                             & 0.67                   & \textbf{0.77}         \\
respiratory hillock cell                        & 0.39                             & 0.78                   & \textbf{0.82}         \\
tracheobronchial smooth muscle cell             & 0.33                             & \textbf{0.78}          & 0.65                  \\
epithelial cell of alveolus of lung             & 0.29                             & 0.18                   & \textbf{0.55}         \\
bronchial goblet cell                           & 0.23                             & 0.04                   & \textbf{0.11}         \\
plasmacytoid dendritic cell                     & 0.18                             & 0.87                   & \textbf{0.94}         \\
acinar cell                                     & 0.15                             & 0.67                   & \textbf{0.81}         \\
lung pericyte                                   & 0.10                             & 0.88                   & \textbf{0.89}         \\
ionocyte                                        & 0.09                             & 0.77                   & \textbf{0.85}         \\
T cell                                          & 0.09                             & \textbf{0.57}          & 0.55                  \\
tracheobronchial serous cell                    & 0.05                             & 0.43                   & \textbf{0.64}         \\
myofibroblast cell                              & 0.04                             & 0.25                   & \textbf{0.69}         \\
conventional dendritic cell                     & 0.04                             & 0.58                   & \textbf{0.81}         \\
mucus secreting cell                            & 0.03                             & 0.61                   & 0.61                  \\
dendritic cell                                  & 0.02                             & 0.46                   & \textbf{0.70}         \\
mesothelial cell                                & 0.02                             & 0.91                   & \textbf{1.00}         \\
smooth muscle cell                              & 0.02                             & 0.09                   & \textbf{0.15}         \\
lung neuroendocrine cell                        & 0.02                             & 0.97                   & 0.97                  \\
brush cell of tracheobronchial tree             & 0.01                             & 0.21                   & \textbf{0.37}         \\
stromal cell                                    & 0.01                             & 0.35                   & \textbf{0.88}         \\
fibroblast                                      & 0.01                             & 0.30                   & \textbf{0.60}         \\
hematopoietic stem cell                         & 0.01                             & 0.78                   & \textbf{0.89}         \\
tracheobronchial goblet cell                    & 0.01                             & 0.00                   & \textbf{0.50}         \\ \hline
\end{tabular}}
\end{table}
\newpage

\subsection{Missing gene imputation with CFGen}
In the scVI paper \citep{Lopez2018}, 10\% of data entries are masked and set to zero, with the model trained on this corrupted data. During inference, masked cells are passed through the encoder, and latent codes $\mathbf{z} \sim q_\psi(\cdot|\mathbf{x})$ are sampled from the posterior. The VAE, trained to handle noisy inputs, decodes $\mathbf{z}$ to infer masked counts. Similarly, we propose an imputation strategy using CFGen as follows: 

\begin{itemize}
    \item Train CFGen on noisy data. 
    \item Encode a noisy input $\mathbf{x}$ into the latent representation $\mathbf{z}_1 = f_\psi(\mathbf{x})$.
    \item Invert the generative flow to compute $\mathbf{z}_0 = \phi_0(\mathbf{z}_1)$, mapping $\mathbf{z}_1$ to noise.  
    \item Sample around $\mathbf{z}_0$ as $\mathbf{z}_0' \sim \mathcal{N}(\mathbf{z}_0, \sigma^2 \mathbf{I})$.  
    \item Transport $\mathbf{z}_0'$ back to $\mathbf{z}_1' = \phi_1(\mathbf{z}_0')$, then decode to impute gene values for $\mathbf{x}$.  
\end{itemize} 

\begin{figure}[H]
\centering
\includegraphics[width=0.7\textwidth]{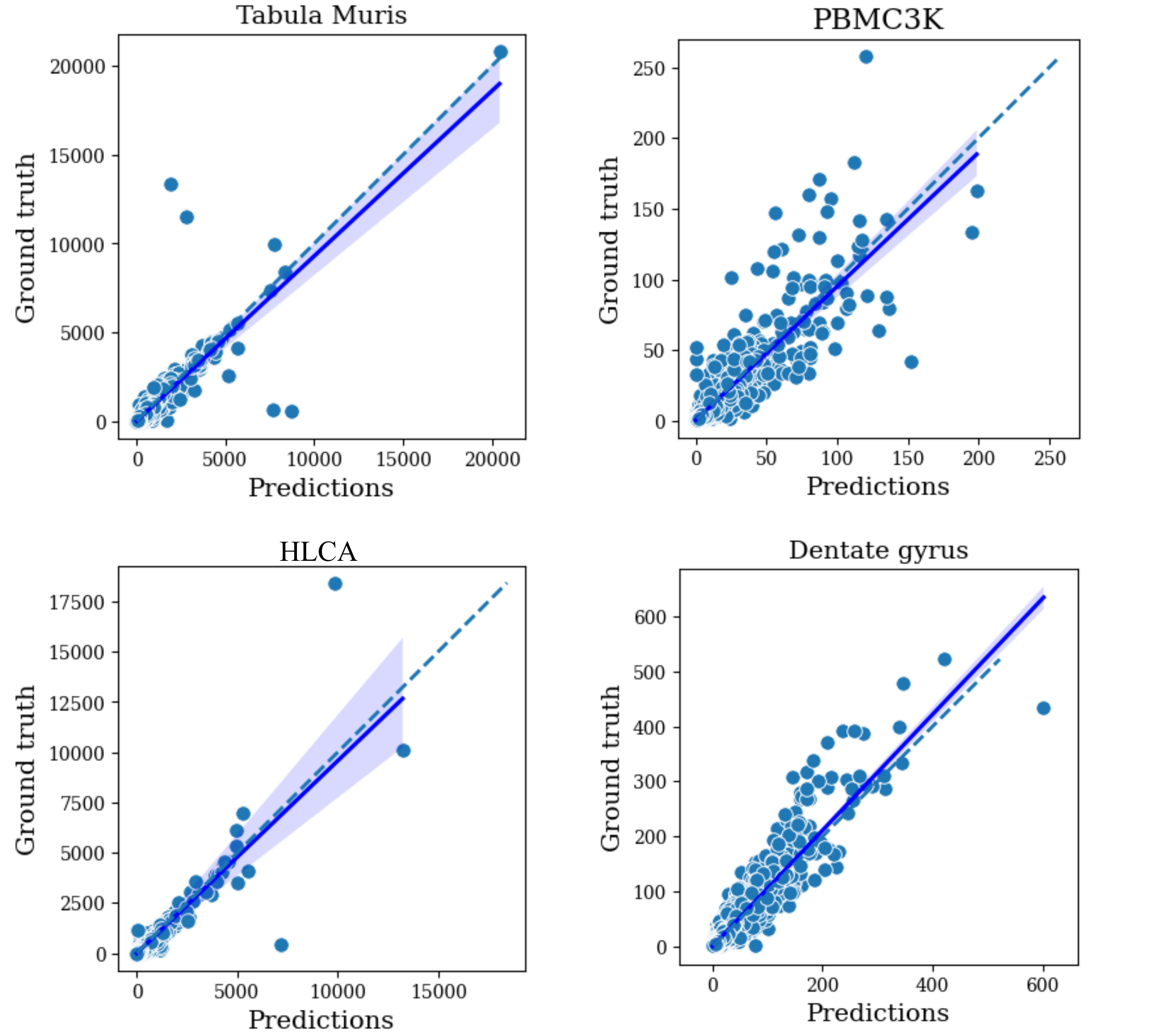}
\caption{Scatterplot between imputed and real gene expression values before masking across datasets. Correlations can be found in \cref{tab:imput_perf}.}
\label{fig: imputation_results}
\end{figure}

We tested this strategy on four datasets, masking 10\% of the counts. \cref{fig: imputation_results} shows that our predictions correlate with pre-masking data, and \cref{tab:imput_perf} demonstrates superior imputation accuracy compared to scVI in three out of four datasets (Pearson correlation, mean absolute distance).

\begin{table}[H]
\centering
\caption{Mean distance and correlation between real and imputed genes by scVI and CFGen.}
\label{tab:imput_perf}
\resizebox{0.90\textwidth}{!}{%
\begin{tabular}{ccccccccc}
\hline
      & \multicolumn{4}{c}{Mean L$_1$ distance real-imputed counts ($\downarrow$)}                            & \multicolumn{4}{c}{Pearson correlation real-imputed counts ($\uparrow$)} \\ \hline
      & PBMC3K        & Dentate gyrus      & HLCA          & \multicolumn{1}{c|}{T. Muris}      & PBMC3K        & Dentate gyrus       & HLCA          & T. Muris      \\ \hline
CFGen & \textbf{1.21} & 0.42          & \textbf{3.21} & \multicolumn{1}{c|}{\textbf{4.81}} & \textbf{0.68} & 0.56          & \textbf{0.83} & \textbf{0.86} \\
scVI  & 1.47          & \textbf{0.35} & 4.43          & \multicolumn{1}{c|}{6.08}          & 0.61          & \textbf{0.58} & 0.75          & 0.79          \\ \hline
\end{tabular}}
\end{table}

\begin{figure}[H]
\centering
\includegraphics[width=1\textwidth]{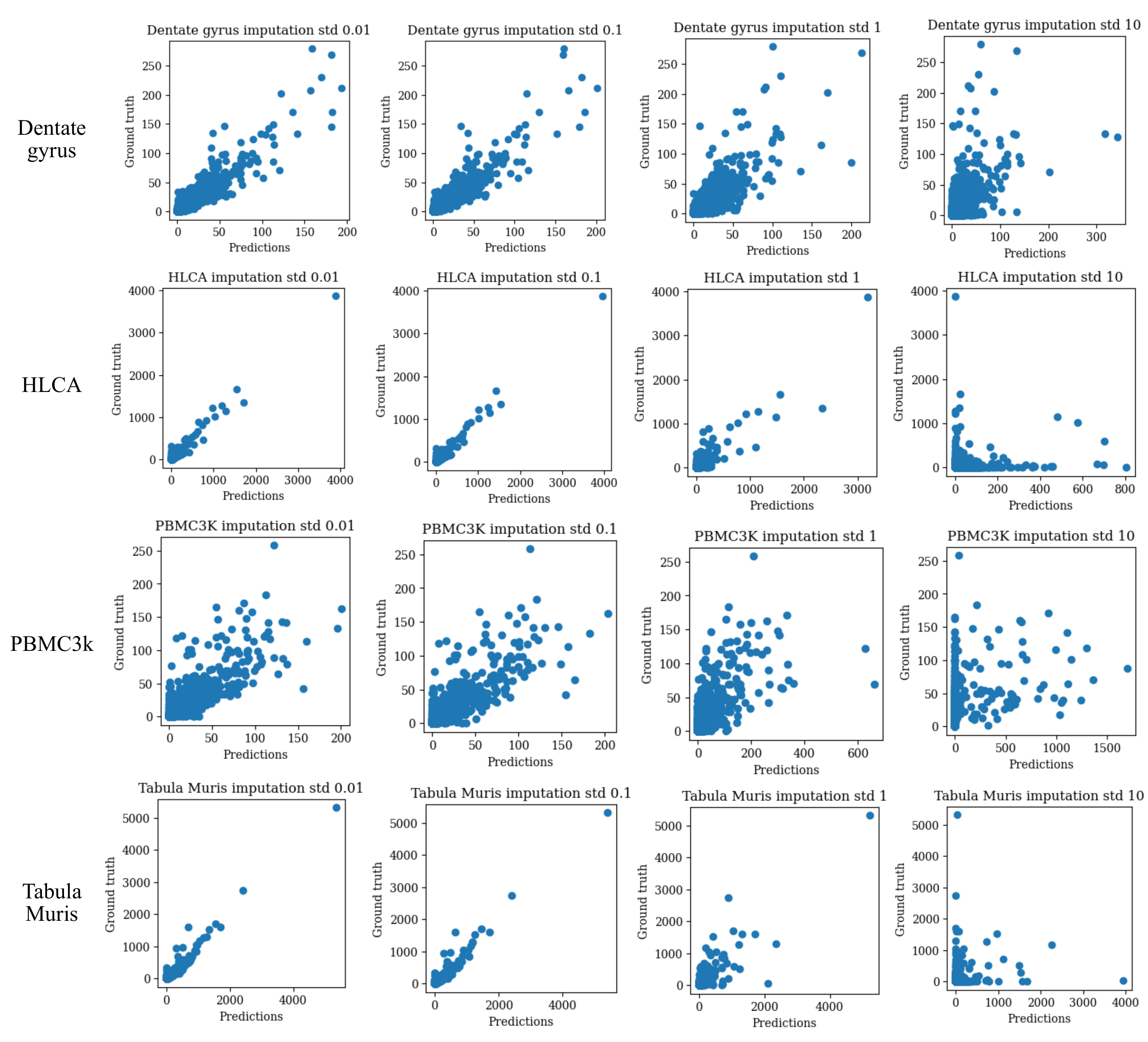}
\caption{Correlation between the CFGen-imputed and real gene expression before masking as a function of the variance of the added noise. Rows represent different datasets, columns stand for the standard deviation of the noise used to sample around the latent representation of the cell. Perfect correlation along the bisector is the best possible imputation result.}
\label{fig: imputation_noise}
\end{figure}

In \cref{fig: imputation_noise}, we study how the quality of the imputation by CFGen varies as a function of the amount of noise used to sample around $\z_0$. Notably, a higher noise leads to worse imputation results, since the generative modeling aspect takes over and samples a completely new cell which loses the structure of the originally encoded noisy observation. Specifically, \cref{fig: imputation_noise} highlights that $\sigma$ should remain below 0.1 to avoid sampling distant $\mathbf{z}_0'$ values, which generate unrelated cells and disrupt correlations with original gene expressions. 

\newpage

\subsection{Additional results multi-label generation}\label{sec: multilab_res}

\begin{table}[H]
\centering
\caption{Extension to \cref{fig: multi_label_guidance}. We train a 3-layer MLP with softmax head on the real data to predict the classes of the two attributes considered for each dataset. For different levels of the combination of guidance weights, the classifier is applied to predict the average probability that the generated observations are of a certain guidance class. When guided on a single attribute it is expected that generated cells are assigned with high probability only to the class of such an attribute. As guidance strength increases for the counterpart attribute, CFGen models the intersections between attributes increasingly better and, therefore, enables high classification probability for both guiding labels.}
\label{tab:my-table}
\resizebox{0.85\textwidth}{!}{%
\begin{tabular}{ccc|ccc}
\multicolumn{3}{c|}{NeurIPS}                                                                                                   & \multicolumn{3}{c}{Tabula Muris}                                                                                               \\
Weights                                                                                         & $p(\mathrm{CD14+ M.})$ & $p(\mathrm{donor \: 1})$ & Weights                                                                                           & $p(\mathrm{Tongue})$ & $p(\textrm{18-M-52})$ \\ \hline
\begin{tabular}[c]{@{}c@{}}$\omega_\mathrm{donor} = 0$\\ $\omega_\mathrm{cell \: type} = 1$\end{tabular} & 0.98          & 0.40         & \begin{tabular}[c]{@{}c@{}}$\omega_\mathrm{mouse \: ID} = 0.0$\\ $\omega_\mathrm{tissue} = 1$\end{tabular} & 0.98        & 0.19         \\ \hline
\begin{tabular}[c]{@{}c@{}}$\omega_\mathrm{donor} = 1$\\ $\omega_\mathrm{cell \: type} = 1$\end{tabular} & 0.96          & 0.87         & \begin{tabular}[c]{@{}c@{}}$\omega_\mathrm{mouse \:ID} = 1$\\ $\omega_\mathrm{tissue} = 1$\end{tabular}   & 0.98        & 0.69         \\ \hline
\begin{tabular}[c]{@{}c@{}}$\omega_\mathrm{donor} = 5$\\ $\omega_\mathrm{cell \: type} = 1$\end{tabular} & 0.96          & 1.00         & \begin{tabular}[c]{@{}c@{}}$\omega_\mathrm{mouse \: ID} = 2.5$\\ $\omega_\mathrm{tissue} = 1$\end{tabular}  & 0.96        & 0.97         \\ \hline
\end{tabular}}
\end{table}

\begin{figure}[H]
\centering
\includegraphics[width=0.8\textwidth]{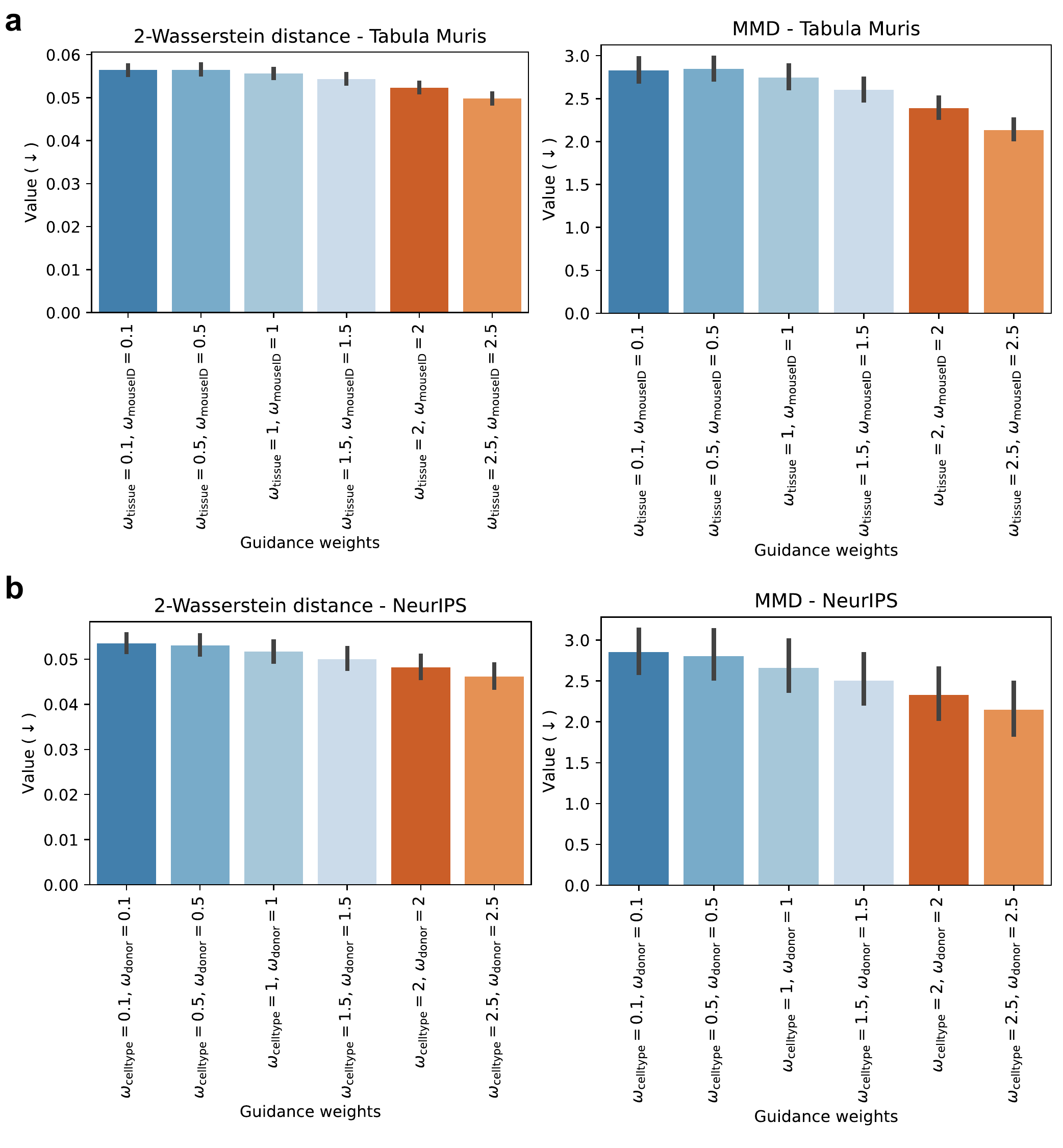}
\caption{Performance on the generation of the intersection of attributes on the Tabula Muris (top)
and NeurIPS (bottom) datasets based on distributional metrics. On the x-axis, we increase the guidance parameters for both conditioning attributes.}
\label{fig: intersection_perf}
\end{figure}
\newpage

\subsection{Additional results on multi-attribute guidance results}\label{app:guidance_selection}
In batch correction, cells are transported to noise and then back to data, guided by both a biological and a target batch covariate. The guidance strength parameters $\omega_{\text{bio}}$ and $\omega_{\text{batch}}$ control the emphasis on biological conservation and batch correction, respectively.

\begin{table}[H]
\centering
\caption{The average batch correction and bio conservation metrics from the scIB package evaluated at different levels of guidance strength.}
\label{tab: metrics_vs_guidance}
\resizebox{0.9\textwidth}{!}{%
\begin{tabular}{c|cc|cc}
\hline
                 & \multicolumn{2}{c|}{C. Elegans}     & \multicolumn{2}{c}{NeurIPS}         \\
Guidance weights & Batch Correction & Bio Conservation & Batch Correction & Bio Conservation \\ \hline
$\omega_{\mathrm{bio}}=$0 $\omega_{\mathrm{batch}}=$0              & 0.48             & 0.55             & 0.32             & 0.63             \\
$\omega_{\mathrm{bio}}=$1 $\omega_{\mathrm{batch}}=$1              & 0.67             & 0.55             & 0.61             & 0.64             \\
$\omega_{\mathrm{bio}}=$1 $\omega_{\mathrm{batch}}=$2              & 0.68             & 0.54             & 0.63             & 0.61            \\
$\omega_{\mathrm{bio}}=$2 $\omega_{\mathrm{batch}}=$1              & 0.68             & 0.63             & 0.64             & 0.73             \\
$\omega_{\mathrm{bio}}=$2 $\omega_{\mathrm{batch}}=$2              & 0.69             & 0.63             & 0.64             & 0.71             \\
$\omega_{\mathrm{bio}}=$2 $\omega_{\mathrm{batch}}=$3              & 0.69             & 0.64             & 0.65             & 0.70             \\
$\omega_{\mathrm{bio}}=$3 $\omega_{\mathrm{batch}}=$2              & 0.70             & 0.67             & 0.65             & 0.77             \\
$\omega_{\mathrm{bio}}=$3 $\omega_{\mathrm{batch}}=$3              & 0.70             & 0.68             & 0.66             & 0.75             \\
$\omega_{\mathrm{bio}}=$3 $\omega_{\mathrm{batch}}=$4              & 0.70             & 0.67             & 0.66             & 0.73             \\
$\omega_{\mathrm{bio}}=$4 $\omega_{\mathrm{batch}}=$4              & 0.70             & 0.69             & 0.67             & 0.77             \\ \hline
\end{tabular}}
\end{table}

\begin{figure}[H]
\centering
\includegraphics[width=1\textwidth]{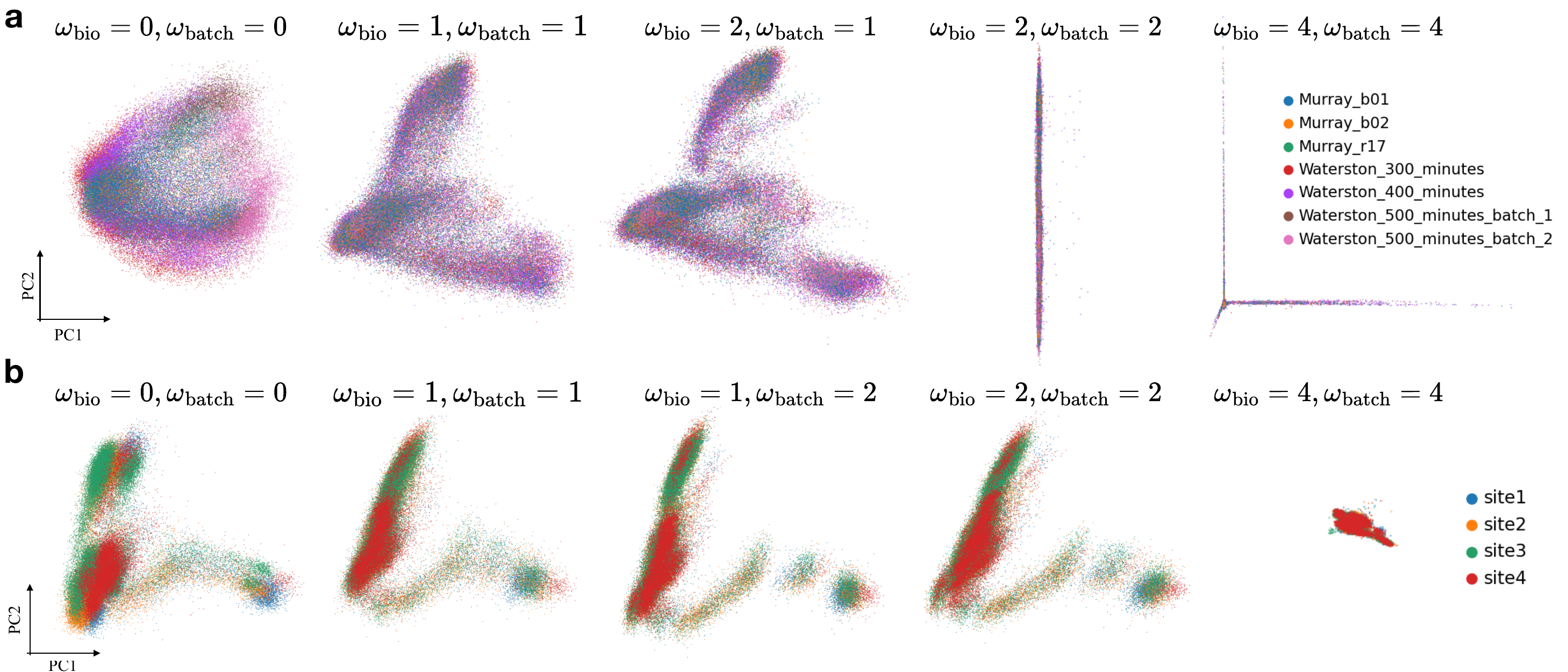}
\caption{The PCA plot of generated cells colored by batch for the C.Elegans (\textbf{a}) and NeurIPS \textbf{(b)} datasets. Each column represents a different combination of guidance strength values.}
\label{fig: corrected_a}
\end{figure}

\begin{figure}[H]
\centering
\includegraphics[width=0.60\textwidth]{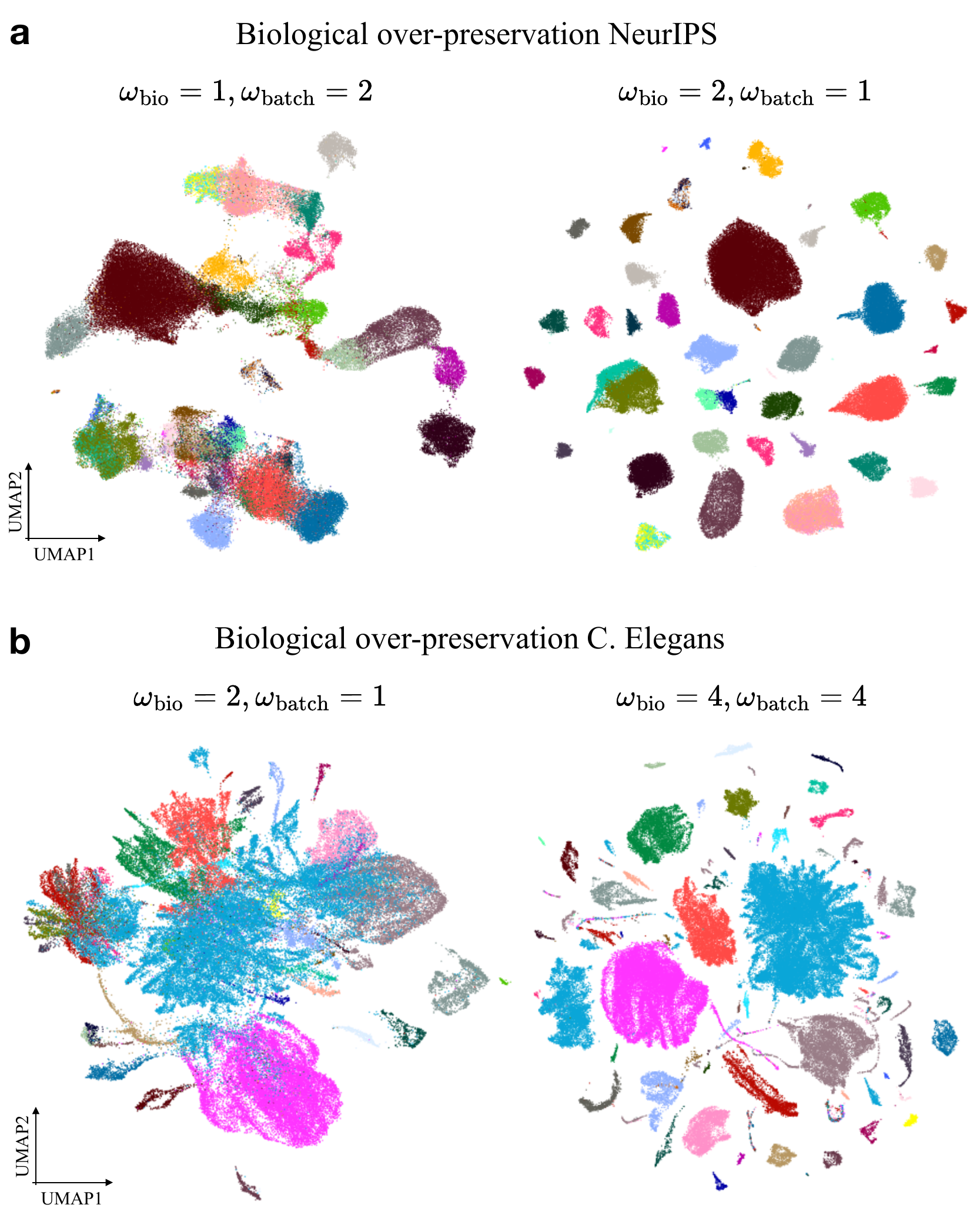}
\caption{The UMAP plot of generated cells colored by batch for the (\textbf{a}) NeurIPS and \textbf{(b)} C.Elegans  datasets. We show one example of generation with a reasonable guidance scheme (left columns) and one with a guidance scheme causing unrealistic cell type distributions (right).}
\label{fig: corrected_b}
\end{figure}

If one relies on the scIB metric in \cref{tab: metrics_vs_guidance}, computed for different guidance strength parameters, the highest possible guidance strengths may appear optimal, as they yield the best aggregation within cell types and batches. However, \cref{fig: corrected_a} and \cref{fig: corrected_b} demonstrate that scIB metrics can be misleading and should be complemented by qualitative evaluation. Excessive guidance in the translation task causes an unnatural collapse of variability beyond what is explained by batch and biological annotations.

We found that guidance strength parameters in the range $\omega_{\text{bio}}, \omega_{\text{batch}} \in \{1, 2\}$ effectively preserve biological signal while performing batch correction without over-squashing cell representations. An example of these unwanted effects is illustrated in \cref{fig: corrected_b}, where excessive biological preservation results in unnatural clustering for both datasets.

Moreover, the severity of batch effects in the data should guide parameter selection. In the C.Elegans dataset, where batch effects are mild, we select $\omega_{\text{bio}} = 2, \omega_{\text{batch}} = 1$ as they provide better performance than $\omega_{\text{bio}} = 1, \omega_{\text{batch}} = 2$ (\cref{tab: metrics_vs_guidance}). Conversely, in the NeurIPS dataset, we observe the opposite effect and thus select $\omega_{\text{bio}} = 1, \omega_{\text{batch}} = 2$. As shown in \cref{fig: corrected_b}, choosing $\omega_{\text{bio}} > 1$ leads to an unnatural biological structure, violating smooth temporal single-cell trajectories.

In conclusion, we recommend first assessing the severity of batch effects in the dataset, and then systematically sweeping over combinations of guidance weights. The optimal configuration should maximize scIB metric values while maintaining realistic single-cell representations.

\subsection{Library size influence in single-cell atlases}\label{sec: library_size_influence}

\end{document}